\documentclass[
  aip,
  jcp,
  amsmath, amssymb,
  reprint
]{revtex4-1}

\usepackage{graphicx}
\usepackage{dcolumn}
\usepackage{bm}
\usepackage[normalem]{ulem}
\usepackage[utf8]{inputenc}
\usepackage[T1]{fontenc}
\usepackage{mathptmx}
\usepackage{amsmath}

\usepackage{color}

\usepackage{soul,xcolor}
\setstcolor{red}

\begin{document}


\title{Low-temperature statistical mechanics of the QuanTizer problem: fast quenching and equilibrium cooling of the three-dimensional Voronoi Liquid}


\author{Tobias M. Hain}
\affiliation{Institut für Mathematik, Universität Potsdam, Karl-Liebknecht-Str.~24-25, D-14476 Potsdam OT Golm}
\affiliation{Physical Chemistry, Center for Chemistry and Chemical Engineering, Lund University, Lund 22100, Sweden}
\affiliation{Murdoch University, College of Science, Health, Engineering and Education, Mathematics and Statistics, 90 South St, Murdoch WA 6150, Australia}
\author{Michael A. Klatt}
\affiliation{Department of Physics, Princeton University, Princeton, NJ 08544, USA}
\affiliation{Present address:
Friedrich-Alexander-Universit\"at Erlangen-N\"urnberg (FAU), Institut f\"ur Theoretische Physik, Staudtstr. 7, 91058 Erlangen, Germany, and
Department of Experimental Physics, Saarland University, Campus E2 9, 66123 Saarbrücken, Germany}
\author{Gerd E. Schr\"oder-Turk}
\affiliation{Murdoch University, College of Science, Health, Engineering and Education, Mathematics and Statistics, 90 South St, Murdoch WA 6150, Australia}
\affiliation{Department of Food Science, University of Copenhagen, Rolighedsvej 26, 1958 Frederiksberg C, Denmark}
\affiliation{Physical Chemistry, Center for Chemistry and Chemical Engineering, Lund University, Lund 22100, Sweden}
\affiliation{Email for correspondence: g.schroeder-turk@murdoch.edu.au}

\newcommand{\highlight}[1]{ {\color{red} \textbf{#1} } }
\newcommand{\ch}[1]{ {\color{blue} \textbf{#1} } }


\date{\today}

\begin{abstract}
The Quantizer problem is a tessellation optimization problem where point 
configurations are identified such that the Voronoi cells minimize the 
second moment of the volume distribution.
While the ground state (optimal state) in 3D is almost certainly 
the body-centered cubic lattice, disordered and effectively hyperuniform 
states with energies very close to the ground state exist that result as 
stable states in an evolution through the geometric Lloyd's algorithm 
[Klatt {\it et al.}~Nat.~Commun., 10, 811 (2019)].
When considered as a statistical mechanics problem at finite 
temperature, the same system has been termed the `Voronoi Liquid' by 
[Ruscher {\it et al.}~EPL 112, 66003 (2015)].
Here we investigate the cooling behaviour of the Voronoi liquid with a 
particular view to the stability of the effectively hyperuniform disordered state.
As a confirmation of the results by Ruscher {\it et al.}, we observe, by 
both molecular dynamics and Monte Carlo simulations, that upon slow 
quasi-static equilibrium cooling, the Voronoi liquid crystallizes from a 
disordered configuration into the body-centered cubic configuration.
By contrast, upon sufficiently fast non-equilibrium cooling (and not 
just in the limit of a maximally fast {\it quench}) the 
Voronoi liquid adopts similar states as the effectively 
hyperuniform inherent structures identified by Klatt {\it et al.}~and 
prevents the ordering transition into a BCC ordered structure.
This result is in line with the geometric intuition that the geometric 
Lloyd's algorithm corresponds to a type of fast quench.
\end{abstract}


\maketitle 



The following article has been accepted by the Journal of Chemical
Physics. After it is published, it will be found at
https://doi.org/10.1063/5.0029301.

\section{Introduction}
\label{sec:intro}


Geometric optimization problems of tessellations search for partitions 
of space into cells with certain optimal geometric properties.
Often these geometric properties can be expressed as energy functionals, 
so that the global optimum corresponds to the ground state of a physical 
system.
Thus, the geometric optimization problem relates fundamental questions 
in mathematics and physics.
Famous examples in three dimensions (3D) are the Kelvin 
problem~\cite{weaire_counter-example_1994,weaire_kelvin_1997} (that is, 
the search for a tessellation with equal volume cells that have 
the least surface area) and the Kepler problem~\cite{hales_proof_2005} 
(that is, the search for cells with the highest packing fraction of 
impenetrable spheres).

While global optima correspond to ground states of physical systems, 
local optima correspond to inherent structures (that is, local minima of 
a complex energy landscapes).
At low-temperature, an equilibrium system will typically be different 
from a non-equilibrium state that is reached by a quench.
Such a fast non-equilibrium cooling often leads to glass-like, highly 
disordered states.
There are, however, also potentials that exhibit disordered ground states 
and that are hence often highly degenerate.
Examples include stealthy potentials, where the ground states suppress 
single scattering up a finite wave number $K$.
The potential suppresses density fluctuations at large 
scales~\cite{torquato_ensemble_2015, zhang_ground_2015-1, 
zhang_ground_2015}, and as a rigorous consequence, the ground state does 
not allow for arbitrarily large 
holes~\cite{zhang_can_2017,ghosh_generalized_2018}.
Another class of examples are models of ``perfect glass'', which involve 
up to four-body interactions and do not allow for crystalline ground 
states (but only disordered configurations)~\cite{zhang_perfect_2016}.
Disordered ground states have also been empirically found in vertex 
models that optimize the isoperimetric ratios of cells (modelling 
biological 
tissues)~\cite{bi_density-independent_2015,merkel_geometrically_2018}.

More generally, we are here interested in models with energy landscapes 
that allow for metastable amorphous states that have energies close to 
the crystalline ground state.

The geometric optimization problem that we consider here is the
\textit{Quantizer
  problem}~\cite{gersho_asymptotically_1979,lloyd_least_1982,conway_sphere_1999,okabe_spatial_2000,du_optimal_2005,du_advances_2010},
that is, we optimize the second moment of inertia of each cell.
Intuitively speaking, the optimization prefers equal-volume cells with
``sphere-like'' shapes. It is a prominent problem in computer science,
where it is applied for example in compression
algorithms\cite{1162229} or mesh generation of two-dimensional
manifolds \cite{6143938}.  In recent years, the Quantizer problem has
attracted attention in physics since it relates to a many-body
interaction that results in surprising physical and geometrical
properties~\cite{torquato_reformulation_2010, ruscher_voronoi_2015,
  ruscher_voronoi_2017, klatt_universal_2019}.

Given a configuration of $N$ points in Euclidean space, the Voronoi 
quantizer (or Voronoi tessellation) assigns to each point $\textbf{z}_i$ 
a cell $C_i$ that contains all sites in space that are closer to that 
point than to any other point in point 
pattern~\cite{okabe_spatial_2000}.
The cells subdivide space without overlap.
The Quantizer energy $E_i$ of a single cell $C_i$ is then defined as the 
moment of inertia of the cell interpreted as a solid object and measured 
with respect to the Voronoi center 
$\textbf{z}_i$~\cite{du_advances_2010, torquato_reformulation_2010,  
ruscher_voronoi_2015}.
More precisely, the total energy $E$ of the system is the sum of the 
cell energies $E_i$, defined as follows: 
\begin{equation}
  \label{eq:quantizer-energy}
  E=\sum_{i=1}^N E_i\quad \text{with}\quad E_i = \frac{\gamma}{2} \int_{C_i} \| 
  \textbf{x} -\textbf{z}_i \|^2 d\textbf{x}.
\end{equation}
where $\gamma$ is a coefficient to set the dimension of 
eq.~\ref{eq:quantizer-energy} to an energy and $N$ is the number of 
points (respectively cells).

The Quantizer problem is defined for a fixed number of points $N$.
Note that this formulation of the Quantizer energy is extensive, in 
contrast to the intensive \textit{Quantizer error}, which is the 
rescaled sum of all single cell energies (see 
\cite{klatt_universal_2019} for more details).
The energy functional can also be expressed by the Minkowski 
tensors~\cite{schroder-turk_minkowski_2011} of the cell $C_i$ (see 
below)~\cite{klatt_universal_2019}.
The Quantizer energy functional assigns an energy to each point 
configuration in Euclidean space.
The Quantizer problem in computer science has thus been reformulated as 
a ground state problem in statistical 
physics~\cite{torquato_reformulation_2010}.

In 3D, the conjectured optimal solution of the Quantizer problem, that 
is, the ground state, is the body-centered cubic (BCC) 
lattice~\cite{conway_sphere_1999}.
It is closely related Kelvin's conjectured equal-volume cells with least 
surface area~\cite{thomson_division_1887,weaire_kelvin_1997}.
A conjecture that was later disproven by the counterexample of Weaire 
and Phelan~\cite{weaire_counter-example_1994}.
The proof of the Kepler conjecture (that no packing of monodisperse 
spheres has a larger packing fraction than the face-centered cubic (FCC) 
packing) reformulates the problem as an optimization problem of 
tessellations (including Voronoi cells)~\cite{hales_proof_2005}.

Here, we are interested in disordered inherent structures with energies 
close to that of the ground state.
Following the approach form \citet{ruscher_voronoi_2015}, we study both 
equilibrium and non-equilibrium dynamics of a many-particle system whose 
energy is defined by a rescaled Quantizer energy, that is, a 
geometrically driven particle system with many-body interactions.
Our focus is on order/disorder transitions, that is, on the 
degree of structural order and disorder of different states and how it 
changes during melting, slow cooling, or a quench.

Ruscher \textit{et al.}~studied in detail this many-particle system
from a thermodynamic point of view~\cite{ruscher_voronoi_2015,
  ruscher_voronoi_2017, ruscher_voronoi_2018} and found intriguing
physical behavior like an anomalous sound
attenuation~\cite{ruscher_anomalous_2017}.  They named the system the
\textit{Voronoi liquid}.  The distinct difference to well-established
model systems such as the Lennard-Jones fluid is that the interactions
in the Voronoi liquid are intrinsically many-body.
\citet{ruscher_voronoi_2015} report theoretical considerations as well
as molecular dynamics (MD) simulations that show that the Voronoi
liquid in many ways behave similar to an ordinary fluid, including a
scaling law for its free energy and derivated quantities as well as
dynamic and structural properties. Furthermore a melting and freezing
transition when heating and cooling the system are found, showing a
metastable state with a hysteresis and under and overcooled states
\cite{ruscher_voronoi_2017}. Ruscher {\it et al.}  studied the Voronoi
liquid as a model glass former where crystallization is prevented by
the integration of a term corresponding to
bidispersity\cite{ruscher_voronoi_2018, ruscher_glassy_2020}. Their
work without the polydispersity term\cite{ruscher_voronoi_2015}, and
the results of this paper, show that the system without polydispersity
and defined by eq.~\ref{eq:quantizer-energy} shows a conventional
order/disorder transition upon heating or cooling.

Here, we are specifically interested in a better understanding of the 
inherent structures of the Quantizer problem.
In a recent study~\citet{klatt_universal_2019}, the so-called Lloyd's 
algorithm~\cite{lloyd_least_1982} was applied to a broad range of 
distinctly different random point patterns to find minimal energy 
point-configurations.
At each step of the algorithm, each point is replaced by the center of 
mass of its Voronoi cell.
\citet{klatt_universal_2019} showed that upon application of a 
sufficient number of iterations of the Lloyd's algorithm all
initial random point configurations converged to configurations that are 
amorphous and universal with the same two-point statistics and Minkowski 
tensors within error bars.
Moreover, these final configurations are effectively hyperuniform, that 
is, they exhibit a strong suppression of large-scale density 
fluctuations~\cite{torquato_local_2003, ghosh_fluctuations_2017, torquato_hyperuniform_2018},
as measured by the hyperuniformity index 
$H=\lim_{k\rightarrow 0} S(k)/\max S(k)$~\cite{atkinson_critical_2016, 
torquato_hyperuniform_2018}
with values of $H\lesssim 10^{-4}$.
We will here refer to these configurations as the \textit{converged Lloyd state(s)}. 

In this article on the Quantizer energy, we study both the equilibrium 
behavior and non-equilibrium quenches.
We thus reproduce and confirm results found by 
\citet{ruscher_voronoi_2015}.
Therefore, we use besides MD also Monte-Carlo (MC) simulations.
Moreover, we vary cooling rates to study both crystallization and the 
freezing in inherent structures.
We thus further probe the energy landscape of the Quantizer problem to 
address the question of the stability of the disordered, effectively 
hyperuniform states to which the Lloyd's algorithm 
converges~\cite{klatt_universal_2019}.
Since we here study in detail the Quantizer problem at finite 
temperature $T$, we refer to it as the ``QuanTizer problem''.

This article is structured as followed: in section \ref{sec:methods}
we give details about our simulations, in section \ref{sec:results} we 
present our results and address the question if a disordered, stable 
state for the Quantizer problem exists, before we give a summary of this 
article in section \ref{sec:conclusion}.

\section{Methods}
\label{sec:methods}

Three different numerical methods for the evolution of a point pattern
are used in this study: Lloyd's algorithm~\cite{lloyd_least_1982} is a
purely geometric algorithm used to compute gradient-descent-like
quenches as previously described and is used in the same way as in
\citet{klatt_universal_2019}. Molecular Dynamics (MD) and Monte Carlo
(MC) codes are used to determine statistical properties of quasistatic
(slowly cooled or heated) systems. Molecular Dynamics is also used to
calculate the non-equilibrium evolution of the system when it is
quenched, that is, with fast cooling rates.

Throughout this article we will use reduced units.
The unit of length is $\lambda=\rho^{-1/3}$, where $\rho$ is the number 
density.
Thus, we choose $\rho=1$.
Each sample contains $N=2000$ particles in a cubic simulation box (of 
side length $2000^{1/3}\lambda\approx 12.6\lambda$) using periodic boundary conditions.
The unit of energy is $[E]=\epsilon=\gamma\lambda^5/1000$, where the factor 
1000 is chosen following \citet{ruscher_voronoi_2015}.
The unit of temperature is $[T]=\epsilon/k$, where $k$ is the Boltzmann 
factor.
All particles have the same mass $m$, which here defines the unit of 
mass.
The arbitrary unit of time for the MD simulation is $[t]=\delta$ 
(Note that the MC simulation and Lloyd's algorithm have no time scale.)

{\bf Monte Carlo method} We use a simple single-step Metropolis algorithm
\cite{metropolis_equation_1953} implemented in the software package
\textsc{Mocasinns} \cite{kruger_mocasinns_2016, Krueger2016}: a trial move is
chosen by selecting a random particle $\textbf{x}_i$ in the system and
move it by a random displacement vector $\Delta \textbf{x}$. The
energy difference $\Delta E$ for this potential new configuration is
then computed. The probability of accepting this trial move $p(\Delta
E)$ is then given by
\begin{equation}
  \label{eq:mc_acceptance}
  p(\Delta E) = \left\{
  \begin{array}{lcl}
    1& \text{for} & \Delta E < 0\\
    \exp \left( -\frac{\Delta E}{kT} \right) & \text{for} & \Delta E \geq 0 \\
  \end{array}
  \right.
\end{equation}
for a given system temperature $kT$. If the trial move is accepted,
the particle is left at its new position, if the move is declined it
is moved back to its original position. The direction of the random
displacement vector $\Delta \textbf{x}$ is random, its length
$\|\textbf{x}\|$ is chosen such as about half of the trial steps are
accepted. This is achieved by checking the acceptance ratio in fixed
intervals of Monte Carlo steps, and doubling the step size if the
acceptance ratio is higher than $0.65$ or cutting it in half for an
acceptance ratio smaller than $0.35$. A lookup table was created to
quickly find an appropriate step size for a given particle number and
temperature.

An essential part of this algorithm is the computation of the energy
difference. To improve performance, only the energies of the
cells affected by the move are recomputed according to
eq.~(\ref{eq:quantizer-energy}). This definition of the energy is
essentially the second moment of the mass distribution of the cell and
thus can be expressed as the trace of the Minkowski Tensor $W_0^{2,0}$:
$E=tr \left[ W_0^{2,0} ( C_i ) \right]$. Minkowski Tensors are a
comprehensive set of metrics, describing geometric properties of a
body \cite{schroder-turk_minkowski_2013,schroder-turk_minkowski_2011}.

The computation of the cell energy is carried out in two steps: first
the Voronoi cell of a particle is computed using the \textit{voro++}
software package \cite{rycroft_voro_2009-1}, the coordinates of the
vertices and edges are then parsed into \textsc{karambola}
\cite{schroder-turk_minkowski_2013}, a tool to compute the
Minkowski tensors. The cell energy is then obtained by computing
$E=tr\left[ W_0^{2,0} ( C_i ) \right]$.  The total energy of a system
is just the sum of all individual cell energies.

\textbf{Molecular Dynamics} is a method to simulate particle systems
by forward integration of Newton's equation of motion $\textbf{a}_i =
\frac{\textbf{F}_i}{m_i}$ in time for each particle, thus computing
the exact trajectory for each constituent. The essential part of each
MD code is thus the computation of the force acting on each
particle. We follow \citet{ruscher_voronoi_2015} and define the force
on the $i$-th particle as $\textbf{F}_i=\gamma \text{\boldmath
  $\tau$}_i$, where $\text{\boldmath$\tau$}_i = V_i \cdot \left(
\textbf{c}_i - \textbf{z}_i \right)$ is the so-called polarisation
vector with $\textbf{c}_i$ being the centroid, $\textbf{z}_i$ the
generator, thus position of the $i$-th particle, and $V_i$ the volume
of the $i$-th cell. The computation of the position of the cell's
centroid as well as its volume is done using the software
\textsc{voro++} \cite{rycroft_voro_2009-1}.

A velocity Verlet integrator coupled with a simple Nos\'{e}-Hoover
thermostat \cite{nose_molecular_1983} was used to integrate the
equation of motions. The Nos\'{e}-Hoover thermostat adds an additional
term $\frac{Q}{2}\dot{s}^2 - (f+1)kT_{eq}$ to the Lagrangian of the
system to reproduce configurations from the NVT ensemble at a
temperature $kT$. The variable $s$ is an variable associated with the
thermostat, $f$ is the number of degrees of freedom of the system,
$kT_{eq}$ the temperature of the NVT ensemble and $Q$ determines the
time scale of the temperature fluctuations introduced by the
thermostat. We implemented the formulation by
\citet{martyna_constant_1994}. In each integration step, time is
advanced by a time step $\Delta t$ and the positions $\textbf{x}_i$
and velocities $\dot{\textbf{x}}_i$ of each particle as well as the
thermostat variable $s$ and its derivative $\dot{s}$ are updated
accordingly.

\textbf{Lloyd's algorithm} is a purely geometric algorithm to minimize
the Quantizer energy. It comprises the reposition of a simple step: move
the generator of a cell $\textbf{z}_i$, thus a particle, to the
centroid of its cell $\textbf{c}_i$.

A typical simulation run, either MD or MC, would consist of the
following steps: first, an initial configuration is initialized with a
given particle number $N$ and system size. 

Initial configurations can be generated as perfect BCC crystal or as
an ideal gas, thus each component of each particle is chosen uniformly
random across the simulation box, corresponding to a binomial point
process. Furthermore simulation can be initialized with arbitrary point
configurations read from simple text files, so a previously saved
simulation snapshot can be used as initial configuration.

The next step is to choose a cooling schedule: a temperature step size
$k\Delta T$ as well as a number of temperature steps $N_{kT}$ is
set. For each temperature a set amount of MD or MC steps, called
\textit{relax steps} are performed to get the system to thermodynamic
equilibrium. Once these are done, another set of steps, called
\textit{measurement steps} are performed and relevant measurements are
taken, the most important being the energy and $\tau$ order
parameter. After these are done, the system temperature is increased
by $k\Delta T$. This is repeated until the final number of temperature
increments has been added. In our MD simulations, the cooling rate is 
defined as $\sigma = \frac{ k\Delta T}{\text{\#relax steps} \cdot \Delta 
t}$ and thus has the units of energy over time.
For MD or MC quenches, this cooling schedule would simply consist of a 
single temperature $kT=0$.

The measurement used to describe structures in this study are
essentially the Quantizer energy, the Structure factor $S(k)$ and the
$\tau$ order metric~\cite{torquato_ensemble_2015}.
The structure factor is essentially the scatter intensity of a 
structure.
For a single snapshot of particles in a cubic box of length $L$ with 
periodic boundary conditions it is given 
as~\cite{hansen_theory_2013, torquato_hyperuniform_2018}
\begin{equation}
  \label{eq:def-sk}
  S(k) =  \frac{1}{N}\left| \sum_{i=1}^N  e^{ -i \textbf{k} \cdot
  \textbf{x}_i } \right|^2
\end{equation}
where the sum runs over all particles in the system, 
$k = \|\textbf{k}\|$, and 
$\textbf{k}\in\{\frac{2\pi}{L}(h,k,l):(h,k,l)\in\mathbb{Z}^3\}$.

The $\tau$ order metric measures spatial correlations on all length
scales and is defined as
\begin{equation}
  \label{eq:tau-definition}
  \tau := \frac{\omega_d}{\left( 2\pi \right)^d} \int _0 ^{\infty}
  k^{d-1} \left[ S(k) -1 \right] ^2 dk,
\end{equation}
where, $d$ is the dimension, in our case $d=3$, $\omega_d$ is the
surface area of a unit ball in $d$ dimensions ($\omega_3 = 4\pi$)
and $S(k)$ the structure factor.  For a completely disordered,
uncorrelated structure $\tau$ vanishes, while it diverges as soon as
Bragg peaks appear, i.e. especially for systems with a perfect long
range order like crystals. Since this parameter unites a large amount
of information into a single scalar value it is prone to mainly two
errors: small changes in the structure can cause significant change in
magnitude of $\tau$, we estimate this error by computing the standard
error of multiple runs with identical parameters to $\Delta
\tau_{\text{stat}} = 0.1 - 0.15$, depending on the parameters
chosen. Systematic errors caused by the choice of the binning of
$S(k)$ as well as an upper integration cutoff $k_{\text{max}}$ can not
be avoided. By computing the standard deviation of different binnings
of a single system we estimate these systematic errors to $\Delta
\tau_{\text{sys}} = 0.9$. Combining both statistical and systematic
errors, we assume a total error of $\Delta \tau \approx 1$, which is
in line with the previous analysis by \citet{klatt_universal_2019}.

A detailed list of the parameters used to generate the data in this 
article is provided in Appendix \ref{sec:sim-params}.
Unless stated otherwise, temperature is quantified by $kT$ with the 
Boltzmann constant $k$ and has units of energy, $[kT]=\epsilon$.
All time steps $\delta t$ have time unit $[\delta t]=\delta$.

\section{Results}
\label{sec:results}

\begin{figure}
  \includegraphics{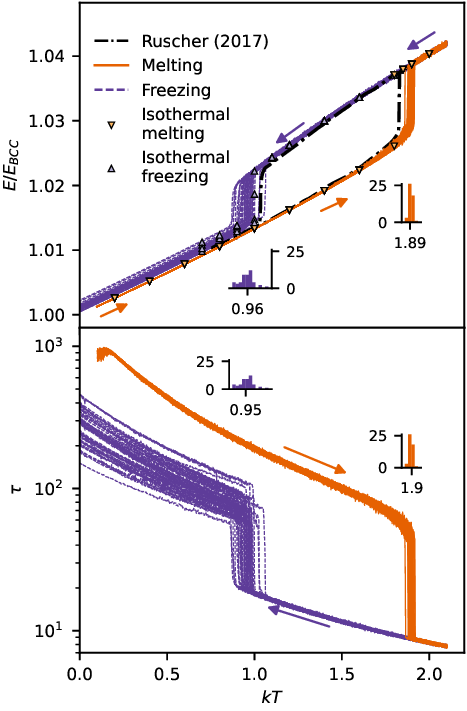}%
\caption{\label{fig:md_phase_transition} Molecular dynamics (MD)
  simulation data of disorder-order (freezing) and order-disorder
  (melting) phase transitions of the Quantizer system observed at
  melting of a BCC crystal and freezing of an ideal gas, and
  comparison to earlier data by \citet{ruscher_voronoi_2017} (black
  dashed-dotted line).  Shown is the energy (as per
  eq.~\ref{eq:quantizer-energy}, top panel) and $\tau$ order metric
  (as per eq.~\ref{eq:tau-definition}, bottom panel) of 48 individual
  runs for melting and freezing processes. The triangles in the top
  panel show the mean energy of isothermal systems run over a long
  time. All runs show a discontinuity in the energy and the $\tau$
  order metric at the transition temperatures that are different for
  freezing and melting and which vary slightly from run to run. The
  insets show the distribution of the transition temperatures.
  Qualitatively the hysteresis between the temperature of the phase
  transition of the disorder/order (freezing) and order/disorder
  (melting) transitions described by \citet{ruscher_voronoi_2017} is
  reproduced. The ordered structure obtained by freezing shows
  variations from the BCC structure, which are most clearly visible in
  $\tau$ and which are indicative of residual defects in the
  structure. See Appendix \ref{sec:sim-params} for simulation
  details.}
\end{figure}

\begin{figure}
  \includegraphics{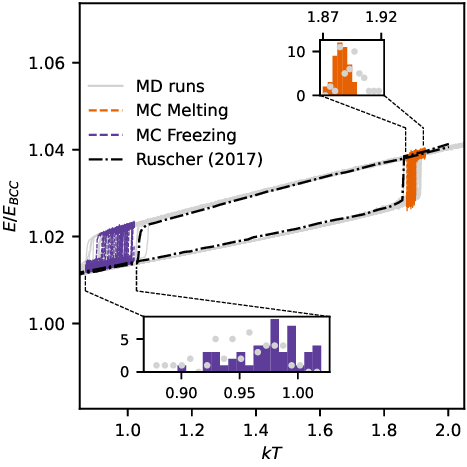}%
\caption{\label{fig:mc_phase_transition} Monte Carlo (MC) simulation 
  data of disorder-order (freezing) and order-disorder (melting) phase 
  transitions of the Quantizer system observed upon heating of a BCC 
  crystal and cooling of an ideal gas.
  The plot shows 48 individual runs for melting and freezing (dashed, 
  colored lines).
  The MC data is consistent with results from MD simulations, see 
  Fig.~\ref{fig:mc_phase_transition}, with slight differences in the 
  variations of the transition temperatures; the MC melting runs show a 
  slightly more narrow distribution of transition temperatures than the 
  MD data, with $kT_{m}=1.886\pm0.006$ and $kT_{f}=0.97\pm0.03$, as seen 
  in the insets showing the distribution of transition temperatures in 
  colored bars and the MD data is indicated by light grey dots.
See Appendix \ref{sec:sim-params} for simulation details.}
\end{figure}

Our key results concern the structures obtained by a quench of the 
system, especially in relation to the converged Lloyd states described by 
\citet{klatt_universal_2019}.
However, we first describe our reproduction of the findings of 
\citet{ruscher_voronoi_2017} of an order/disorder transition encountered 
upon slow equilibrium melting or cooling.
Figures \ref{fig:md_phase_transition} and \ref{fig:mc_phase_transition} 
show our results regarding the order/disorder transition upon slow 
equilibrium cooling or melting.

Figure~\ref{fig:md_phase_transition} shows MD simulations essentially of 
the same system investigated by 
\citet{ruscher_voronoi_2015,ruscher_voronoi_2017} and 
\citet{ruscher_anomalous_2017}. When a Voronoi liquid of $2\times 
10^3=2000$ particles is heated slowly (with a heating rate of 
$\sigma=2.5\times 10^{-4}{\epsilon}/{\delta}$) starting from positions 
on a BCC lattice at $kT=0.1$, it undergoes an order/disorder transition; 
at $kT_{melt}\approx 1.89$, the Voronoi liquid abruptly changes from 
configurations that represent oscillations around the lattice sites to a 
configuration with no apparent order. Upon further heating this 
structure is characterised by the correlations typical for a fluid (for 
very high temperatures we expect it to converge to an ideal gas). This 
transition is evident in the energy $E$ as well as in the structural 
order parameter $\tau$. In our simulations of 48 systems, the transition 
temperatures vary slightly, with an average of $1.89$ and a standard 
deviation of $0.01$. These transition temperatures are close to, but 
slightly above the observed transition $kT\approx1.85$ in 
\cite{ruscher_voronoi_2015,ruscher_voronoi_2017}.
Our curves for the structure factor in the liquid phase agree 
with those from~\cite{ruscher_voronoi_2017}.

Upon cooling (with the same slow rate as above and starting from an 
ideal gas configuration at $kT=2.1$), the system shows the reverse 
transition from a disordered state to an ordered BCC-like state, at a 
temperature $kT_{cool}=0.96\pm0.04$ (again close to but slightly below 
the temperature $kT\approx1.05$ in 
\cite{ruscher_voronoi_2015,ruscher_voronoi_2017}). In line with 
expectation for the hysteretic behaviour of typical first-order ordering 
transitions, the transition back to the ordered BCC-like state occurs at 
a temperature lower than the transition upon heating, i.e., 
$T_{cool}<T_{melt}$. The ordered structures obtained from this cooling 
process are slightly less ordered than those obtained by slowly heating 
up an initially perfect BCC lattice; this is evident in a very slight 
discrepancy in the energy value (which on the interval $kT\in[0.25,0.8]$ 
is on average 0.116\% higher than the energy values); it is even more 
evident in the structure factor which is sensitive to structural detail.
The deviations are quantified by the $\tau$ order metric and shown on a 
logarithmic scale in Fig.~\ref{fig:md_phase_transition}.

The above findings obtained on systems that are cooled down or heated
up are supported by 'isothermal simulations', that is, simulations at
a fixed temperature with significantly longer simulation runs with
$8.5\times 10^5$ MD steps (as opposed to the $10^4$ MD steps per
temperature step in the cooling/melting simulations above).

The mean energies of these isothermal simulations, shown in
fig.~(\ref{fig:md_phase_transition}), support our earlier
results. Four independent realisations of the system are simulated for
each temperature as shown in the figure; data labelled 'isothermal
melting' results from simulating a system from an initial BCC
configuration, whereas 'isothermal freezing' refers to a system
prepared from an initial ideal gas system; for further simulation
details see Appendix \ref{sec:sim-params}. The mean energy values of
the isothermal simulations coincide largely with the energy
trajectories of the slow cooling and heating processes, except in a
small region around the order/disorder and disorder/order
transitions. We thus conclude that the heating and cooling can be
assumed a quasi static process except in the vicinity of the
order/disorder and disorder/order transition.

The transitions (discontinuities) in the mean energy of the isothermal
simulations are slightly offset from the values of the continuous
cooling/heating transitions. The discontinuity of the mean energy of
the isothermal heating systems occurs at a slightly lower temperature
than all of the slowly heated systems; the discontinuity of the mean
energy of the isothermal freezing systems occurs at a slightly higher
value than the average transition temperature of the slowly cooled
systems. This trend could indicate that our slowly cooled or heated
systems are not quite fully equilibrated, but fairly close to
equilibrium.

We note that the freezing isothermal systems show a similar spread of
energy values after the disorder/order transition than the slowly
cooled systems, indicating that the residual defects are not caused by
too little equilibration time.

Our results (both for slow cooling/heating and for the isothermal
simulations) are in good agreement with those from Ruscher's study
\cite{ruscher_voronoi_2017} of the Voronoi liquid.  We add three
comments in regards to the agreement:

(1) Ruscher's value for the melting temperature ($kT\approx 1.85$) is
slightly lower than ours ($kT_{m}\approx 1.886$), and Ruscher's value
for the freezing temperature ($kT_f\approx 1.05$) slightly higher than
ours ($kT\approx 0.96$).  These slight differences are probably due to
our slightly faster cooling or heating rates, to the different system
sizes and to slightly different thermostats and simulation
parameters. (We note that, as expected for slower cooling rates,
Ruscher finds a transition temperature for the isothermal melting
which is slightly closer to the isothermal melting system than
ours. Within this systematic error due to slight differences in
cooling/heating rates, we consider that our results agree with those
of Ruscher.)

(2) At the finite size of our simulations ($N=2000$ particles), we
find a distribution for the values of the freezing temperature and the
cooling temperatures which are (average $\pm$ standard deviation)
$kT_{f}=0.96 \pm 0.04$ and $kT_{m}=1.89\pm0.01$ for the MD simulations
(\ref{fig:md_phase_transition}) and $kT_{freeze}=0.97 \pm 0.03$ and
$kT_{melt}=1.886\pm0.006$ for the MC simulations
(\ref{fig:mc_phase_transition}).  We understand that Ruscher only
presented data for a single simulation run with about 8000 particles.

(3) In our simulations, the final energy of the freezing curves
differs slightly from that of the perfect BCC lattice.  Visual
examination shows that this is due to residual disordered artifacts in
the otherwise ordered lattice.  These deviations are more clearly
visible in the $\tau$ parameter which is sensitive to small deviations
from order.

Figure~\ref{fig:mc_phase_transition} shows Monte Carlo simulation data for 
the same system which are consistent with the Molecular Dynamics 
simulation data shown in Fig.~\ref{fig:md_phase_transition}, thereby 
providing further confirmation of these results. The quantitative 
values for the transition temperatures are consistent (within error 
bars) in MD and MC simulations; however, we observe a narrower spread of 
transition temperatures for the melting process in our MC simulations as 
compared to our MD simulations, see 
Tab.~\ref{tab:transition-temperatures}.

\begin{table}
  \begin{tabular}{l|c|c}
    \hline
    & $kT_{f}$ & $kT_{m}$\\
    \hline
    \hline
    Molecular dynamics & $0.96 \pm 0.04$ & $1.89\pm0.01$\\
    Monte Carlo & $0.97\pm0.03$ & $1.886\pm0.006$ \\
    \citet{ruscher_voronoi_2017} \hspace*{1.5cm}&\hspace*{0.9cm}  $1.05$   \hspace*{0.9cm}&\hspace*{0.9cm}   $1.85$     \hspace*{0.9cm}\\
    \hline
  \end{tabular}
  \caption{\label{tab:transition-temperatures} Transition temperatures 
  $T_m$ for the order/disorder transition (melting) and $T_f$ for the 
  disorder/order (freezing) transition, as computed by molecular 
  dynamics simulations, Monte Carlo simulations and as reported by 
  \citet{ruscher_voronoi_2017}. The notation is $T\pm \delta T$ where 
  $T$ is the average over all simulation runs and $\delta T$ the 
  standard deviation. The values for Ruscher's data are estimates 
  extracted from diagrams in her article\cite{ruscher_voronoi_2017}.}
\end{table}

This concludes our analysis of the quasi-static (slow) equilibrium 
cooling and heating of the Voronoi liquid. Confirming the results by 
Ruscher {\it et al.} we find the system to behave like a typical 
first-order order/disorder transition with hysteresis in that limit, 
with the low-temperature state given by the BCC lattice. 

We now turn to a different question, namely that of what structures the 
Voronoi liquid adopts upon fast non-equilibrium cooling or a quench.
These data are obtained by MD simulations where the system is 
initialized in equilibrium configurations at high $T$, and then cooled 
at high cooling rates.
The limit of an infinite cooling rate, where the temperature is 
abruptly set to $0$, is here referred to as {\it quench}.
These non-equilibrium final configurations are compared in 
particular to the structures obtained by Lloyd's algorithm, discussed by 
\citet{klatt_universal_2019}; Lloyd's algorithm represents a 
steepest-descent minimization of the energy functional in 
eq.~\ref{eq:quantizer-energy} and can therefore be regarded as a 
type of quench.

\begin{figure}
\includegraphics{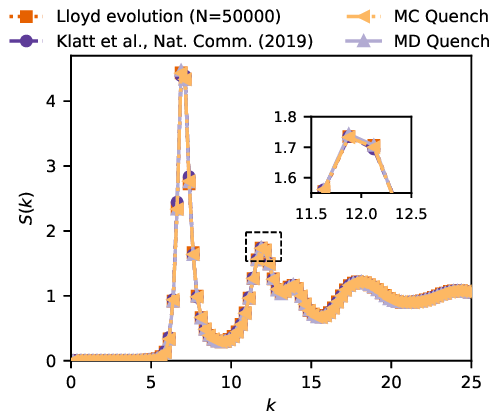}%
\caption{\label{fig:quench} Structure factor $S(k)$ of a quenched
  ideal gas (binomial point-process) at $kT=0$ with $N=2000$ particles
  and a particle density $\rho=1$ using Molecular dynamics (MD), Monte
  Carlo (MC) and the gradient-descent-like Lloyd's algorithm. Each
  structure factor is averaged over 24 individual runs. A detailed
  description of the processes and parameters are given in the method
  section \ref{sec:methods}. All methods evolve into disordered
  structures with energies and values of the $\tau$ order metric equal 
  within
  measurement uncertainties, see tab.~(\ref{tab:energies}). According
  to our data, these structures are identical to the universal,
  amorphous structures previously found by \citet{klatt_universal_2019}
  as remarkably stable, disordered minimal energy configurations of
  the Quantizer system. See Appendix \ref{sec:sim-params} for simulation details.}
\end{figure}

Figure~\ref{fig:quench} and Table~\ref{tab:energies} represent our
analysis of the structure of the configurations that result from
quenching the system, that is, by MD or MC simulations of a systems
where a high-temperature ideal gas configuration evolves when the
temperature is abruptly set to 0. Figure~(\ref{fig:quench}) shows the
structure factors of the structures thus obtained and Table
\ref{tab:energies} the $\tau$ order metric calculated from these. See
also Appendix \ref{sec:lloyd-vs-md} for a technical comparison of an
MD quench and Lloyd's algorithm.

These final structures are compared to the structures obtained by
application of the (purely geometric) Lloyd's algorithm to the same
structures, as suggested by \citet{klatt_universal_2019} (and also to
the data for that same system as reported by
\citet{klatt_universal_2019}).  The key result is that, to a high
degree of accuracy and within the statistical accuracy of our data,
structures obtained by MD or MC quenches are indistinguishable from
the converged Lloyd states in terms of the structure factor and the
derived $\tau$ metric (within error bars).

Structural metrics for local order and local packing structure, namely the Minkowski structure metrics and cell statistics of the Voronoi tessellation, also
show good agreement of the final structures of the different
quenches. See Appendix \ref{sec:characterisation} for further details.
Moreover, we also find a hyperuniformity index $H$~\footnote{We
  estimate $H$ by fitting a Gaussian-shaped density to the peak
  ($6<k<7.5$) and by extrapolating a second-degree polynomial $ax^2+c$
  to the origin ($k_{\min}<k<4$).  The precise value of $H$ depends on
  the fit range and functional form.}  of the order of magnitude
$10^{-4}$.

Up to here, we have investigated the two extreme cases being (a) slow 
quasi-static heating and cooling (which leads to a hysteric 
order/disorder transition) and (b) quenching as the limit of maximally 
fast non-equilibrium cooling. We now turn our attention to the 
intermittent regime of cooling processes which are too fast to be 
quasi-static yet are not a quench.

\begin{figure*}
  \includegraphics{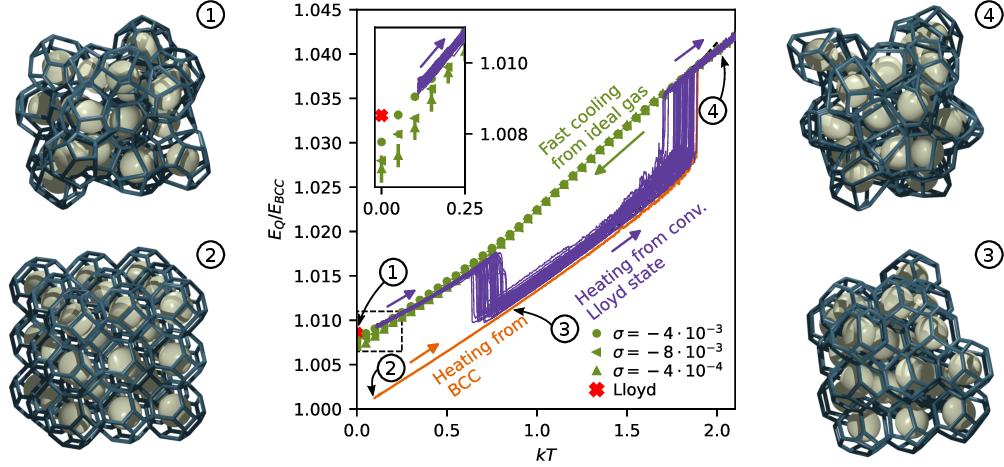}%
\caption{\label{fig:klatt-melting} Non-equilibrium cooling with a fast
  but finite cooling rate and slow equilibrium heating of the
  converged Lloyd states of \citet{klatt_universal_2019}.  The fast
  non-equilibrium cooling of a high-temperature ideal gas results in
  disordered structures very similar to the converged Lloyd
  states~(dotted green curves).  The dependence on the cooling rate
  $\sigma$ is only weak for $|\sigma|>4 \cdot 10^{-4} \epsilon /
  \delta$ (see inset).  Upon cooling, the energy initially follows the
  same functional form as for a slow equilibrium cooling, however it
  fails to show any sign of the ordering transition (which the slowly
  cooled systems undergo for $kT_{melt} \approx 0.96$).  When, from
  small $T$, the converged Lloyd states are heated up slowly at
  $\sigma=2.5\times10^{-4}$ (purple curves), they initially show at
  $kT\approx 0.74\pm0.05$ a transition to a (softened) BCC
  configuration which then melts at a slightly lower phase transition
  temperature at $kT'_{m}\approx 1.80 \pm 0.04$.  The orange curve
  represents the slow equilibrium melting transition starting from a
  low-temperature BCC phase.  All data shown here is obtained by
  Molecular Dynamics simulations.  See Appendix \ref{sec:sim-params}
  for simulation details.  On the left and right hand side, we show
  snapshots taken from MD simulations.  }
\end{figure*}

The key result of Fig.~\ref{fig:klatt-melting} is that, upon rapid
non-equilibrium cooling with sufficiently fast but finite cooling
rates, the system tends to avoid a transition to an ordered (BCC)
structure but instead converges to configurations that are similar in
structure (as measured by the structure factor) and similar in energy
values to the converged Lloyd states.  If the cooling is sufficiently
fast, the value of the cooling rate only has a minor effect on the
final structure that is reached at $kT=0$; the remaining minor
differences in energy are visible in the insert of
Fig.~\ref{fig:klatt-melting} and are significantly smaller than the
difference between the converged Lloyd states ($E/N=118.82\pm0.02$)
and that of the BCC crystal ($E/N=117.815$).

For the fast cooling rates, the majority of the system do not show a
phase transition.  At a cooling rate of $-4 \cdot 10^{-3}$ none of the
24 runs showed a phase transition, out of the 24 runs at a cooling
rate of $-8 \cdot 10^{-3}$ only one underwent a phase transition, and
three out of 24 runs showed a phase transition at a cooling rate of
$-4 \cdot 10^{-4}$.  (These runs are omitted in the data shown here.)

Finally, we have analysed the slow quasi-static heating of a system
that is prepared at $T=0$ in converged Lloyd states.  When heated
slowly, the system energy gradually increases until a certain
temperature following the same (or a very similar) curve to the rapid
cooling cycles, in reverse.  At a certain temperature (the value of
which varies, potentially due to finite size effects), the energy
jumps down to almost the energy of a BCC crystal being melted to the
same temperature.  When heated further, the system behaves similar to
that of the melting curve of a system that started from a BCC crystal.
There are small remnant differences in the energy (the purple curves
in Fig.~\ref{fig:klatt-melting} are slightly above the orange curve)
and the eventual transition to a disordered structure occurs at a
slightly lower temperature.  We do not know the exact nature of these
slight differences.  The degree of residual randomness in these
intermediate BCC-like states appears to facilitate a melting
transition at a slightly lower temperature.

The ``drop'' to a BCC-like states can be avoided, when a system
prepared as a converged Lloyd state at $T=0$ is quickly heated. At a
heating rate of $\sigma \approx 9.3 \cdot 10^{-3}$ all out of 24
individual runs avoid the intermediate BCC-like state and follow the
curve of a quickly cooled system up to the liquid state. With
decreasing heating rates the systems get more likely to fall back into
the intermediate states: at a heating rate of $\sigma \approx 4.3
\cdot 10^{-3}$ 6 out of 24 runs return to the BCC like state. This
behavior while heating is analog to the freezing case: here too the
BCC ground state can be avoided if the system is cooled very quickly.

\begin{table}
  \begin{tabular}{l|c|c}
    \hline
    &$E / E_{BCC}$&$\tau$\\
    \hline
    \hline
    MD Quench&$ 1.00844 \pm 0.00002 $&$33\pm1$\\
    MC Quench&$1.00854\pm0.00003$&$33\pm1$\\
    Lloyd's Iteration\hspace*{1cm}&\hspace*{0.5cm}$1.00852\pm0.00003$\hspace*{0.5cm}&\hspace*{0.5cm}$33\pm1$\hspace*{0.5cm}\\
    \citet{klatt_universal_2019}&$1.008\pm0.001$&$32\pm1$\\
    \hline
  \end{tabular}
  \caption{\label{tab:energies}Final energies and $\tau$ values of
    quenched ideal gas systems as described in
    fig.~(\ref{fig:quench}). The $\tau$ order metric 
    measures the degree of order in the system, it diverges for
    crystalline structures and is zero for complete spatial randomness.
    The value of the $\tau$ order metric from \citet{klatt_universal_2019} is that for
    configurations from Binomial point-processes (from Table~2 in
    the supplementary information); here, we added systematic errors
    discussed both in \citet{klatt_universal_2019} and our method section.
    See Appendix \ref{sec:sim-params} for simulation details.}
\end{table}

\section{Conclusion}
\label{sec:conclusion}

We have studied the configuration of many-particle system that are 
formed by the many-body interaction of the Quantizer 
energy~\cite{du_optimal_2005, du_advances_2010, 
torquato_hyperuniform_2018}, that is, of the Voronoi 
liquid~\cite{ruscher_voronoi_2015}.
We confirmed the freezing and melting transitions found by 
\citet{ruscher_voronoi_2015} using both MD and MC simulations.
A slow cooling in equilibrium leads to the formation of BCC 
crystallites, as expected, since the conjectured ground state (at $T=0$) 
is the BCC lattice.

In contrast, a quench from high temperature states leads to disordered
amorphous structures, more similar to the amorphous states found by
the Lloyd's algorithm~\cite{klatt_universal_2019}. A finite cooling
rate results in final energies slightly below that of the final state
of Lloyd's algorithm, but as we increase the cooling rate the final
energies increase. A quench at $T=0$ leads to final states whose
energies, two-point statistics and local multi-point statistics
coincides within our statistical accuracy with those of the converged
Lloyd states.

To explain both the similarities and differences between a fast MD quench 
and the Lloyd's algorithm, we derive a limit in which a modified MD 
quench, where the mass of a particle is given by the volume of its cell, 
coincides with the iterations of the Lloyd algorithm.

Melting the amorphous converged Lloyd states, we find that the system
remains on the amorphous branch for a finite range of temperatures
(before the system returns to crystalline branch), which agrees with
the meta-stability of the converged Lloyd states.

In future work, MC simulations of the ``QuanTizer problem'' (i.e., Voronoi 
liquid) make it possible to determine the density of states (e.g., using 
the Wang-Landau algorithm~\cite{wang_efficient_2001}) to further study 
the intriguing energy landscape of this many-particle interaction.

Lloyd's algorithm is a gradient-descent minimization method
tailored to the Quantizer problem, in the following sense: The displacement of each point into the direction of the center of mass of its Voronoi cell corresponds to the direction of the negative gradient. Further, the displacement into the center of mass of the Voronoi cell (rather than just in that direction), provides a 'local optimum displacement' for the individual cell. The question arises naturally what final structures
are obtained when applying other energy minimization methods such as the
conjugate gradient descent method, BFGS or others. Preliminary results indicate,
that conjugate gradient methods, steepest descent as well as BFGS
minimization methods as implemented in the
GSL \footnote{See https://www.gnu.org/software/gsl/ for details on the implementation of the minimization methods and further information} seem to evolve random initial structures into amorphous structures that are very similar to the
inherent structure found by Lloyd's algorithm (Note that these algorithms were provided with the gradient direction from Lloyd's algorithm, as a closed formula for the gradient of the energy functional is not available.) A quantitative statistical analysis needs to include a detailed analysis of the observed cases where the minimization methods get stuck in seemingly shallow local minima, and is beyond the scope of this article.  

Ultimately, the research presented in this article supports the search for disordered ground states and 
long-lived inherent structures that offer novel physical properties due 
to their isotropy (in contrast to their crystalline counter parts).
A key question for experimental realizations is the role of long- and 
short-range interaction in such systems.

\section{Acknowledgements}

We are grateful to Massimo Ciamarra who pointed us towards the work of 
Ruscher and colleagues on the Voronoi Liquid.
We thank Salvatore Torquato for discussions and helpful comments,
and we thank J\"org Baschnagel, Jean Farago, and C\'eline Ruscher for 
helpful comments and interesting questions.
This work was supported in part by the Princeton University Innovation 
Fund for New Ideas in the Natural Sciences.
This work was supported by resources provided by the Pawsey 
Supercomputing Centre with funding from the Australian Government and 
the Government of Western Australia, as well as LUNRAC at Lund, Sweden.

\section{Data and Code availability}

The data that support the findings of this study and the code used to
generate them are available from the corresponding author upon request.

\section{Appendix: Algorithmic equivalency, additional structure metrics, and simulation details}

\subsection{Lloyd's algorithm as a limit of an MD quench}
\label{sec:lloyd-vs-md}

We showed that a fast MD quench results in a structure similar to the
converged Lloyd states.
Here we discuss
the conditions under which an MD quench collapses to a ``Lloyd quench''. MD
simulations compute the time evolution of particles, where each step
advances time by an increment $\Delta t$, thus the position of the
$i$-th particle at time $t$, given by $\textbf{r}_i(t)$, is equivalent
to $\textbf{r}_i(n \Delta t) =: \textbf{r}_{i,n}$.

Since Lloyd's algorithm is missing an intrinsic definition of
a time scale, a Lloyd quench can only be compared to an MD quench on a
step by step basis. The position of the $i$-th particle at step $n$
is denoted by $\textbf{r}_{i,n}$. A single Lloyd's iteration is then given
by $\textbf{r}_{i,n+1} = \textbf{c}_{i,n}$, where
$\textbf{c}_{i,n}$ is the centroid of the Voronoi cell associated to
the $i$-th particle at step $n$.

The position of the $i$-th particle after one MD step of time length
$\Delta t$ is given by $\textbf{r}_{i,n+1}=\textbf{r}_i(t+\Delta t)$ which
can be approximated by its Taylor series
\begin{equation}
  \label{eq:taylor-r}
  \textbf{r}_{i,n+1} = \textbf{r}_{i,n} + \dot{\textbf{r}}_{i,n} \Delta t + \frac{\left( \Delta t \right)^2}{2} \ddot{\textbf{r}}_{i,n} + \text{higher orders}
\end{equation}
where a dot denotes the time derivative:
$\dot{\textbf{r}}=\frac{\partial}{\partial t} \textbf{r}$. The force
acting on particle $i$ is given as $\ddot{\textbf{r}}_i =
\frac{\textbf{F}_i}{m_i} = \gamma \frac{V_i}{m_i} \left( \textbf{c}_i
- \textbf{r}_i \right)$. Substitute into eq.~\ref{eq:taylor-r} yields
\begin{equation*}
  \textbf{r}_{i,n} = \left( 1 - \frac{\gamma V_i}{m_i} \frac{ \left( \Delta t \right)^2}{2} \right) \textbf{r}_{i,n} + \left( \frac{\gamma V_i}{m_i} \frac{ \left( \Delta t \right)^2}{2} \right) \textbf{c}_{i,n} + \dot{\textbf{r}}_{i,n}
\end{equation*}
where we neglect orders higher than second derivative. For an MD step being equivalent to a Lloyd iteration $\textbf{r}_i \left(t+\Delta t\right) = \textbf{c}_i$ must hold. Thus, the time step $\Delta t$ must be chosen as
\begin{equation*}
  \Delta t = \sqrt{ \frac{2}{\gamma} \frac{m_i}{V_i} }
\end{equation*}
For this equation to hold, the masses of all particles must be set equal 
their volume before each simulation step. The temperatures are set to 
zero after each simulation step, this acts as a thermostat simulating 
a quench. In this limit an MD step is equivalent to a Lloyd iteration.

On the one hand, this limit demonstrates similarities of Lloyd's 
algorithm and a quench for Voronoi tessellations with energies 
close to the ground state (with a sharp cell volume distribution).
On the other hand, the analysis reveals subtle differences that may lead 
to a slightly different energies and (global) structures.

\subsection{Local structure metrics}
\label{sec:characterisation}

Structures can be locally characterized using the so-called Minkowski
metrics
\citep{schroder-turk_minkowski_2011,schroder-turk_minkowski_2013}. These
define several scalar, vectorial and tensorial quantities measuring
the shape of a convex body $K$ such as a Voronoi cell. Here we
specifically employ the surface Minkwoski tensors $W_1^{0,s}$
describing the radial distribution of outer normal vectors of $K$
\cite{klatt_universal_2019}. It can be conveniently decomposed into
spherical harmonics

\begin{equation*}
  \rho_s^m (K) = \sqrt{ \frac{4\pi}{2s+1}} \frac{\sum_k A_k Y^m_s \left (\textbf{n}_k \right)}{ \sum_k A_k}
\end{equation*}

where $A_k$ are the surface areas of the faces of the body $K$ and
$\textbf{n}_k$ the outer normal vectors. From this tensor rotational
invariants can be constructed:

\begin{equation*}
  q_s \left( K \right) := \sum_{m=-s}^{s} | \rho_s^m (C)|^2
\end{equation*}

These $q_s$ describe the shape of a cell independent of its size and
orientation and thus can be used as a shape metric. Here we compute
the quantities $q_{2,4,5,6,8}$ for each cell in a system and show
their normalized distribution.

The distribution of the number of edges of faces as well as the number
of faces of each cells in the Voronoi tessellation is presented as a
further measurement of the local packing.

\begin{figure}
  \includegraphics{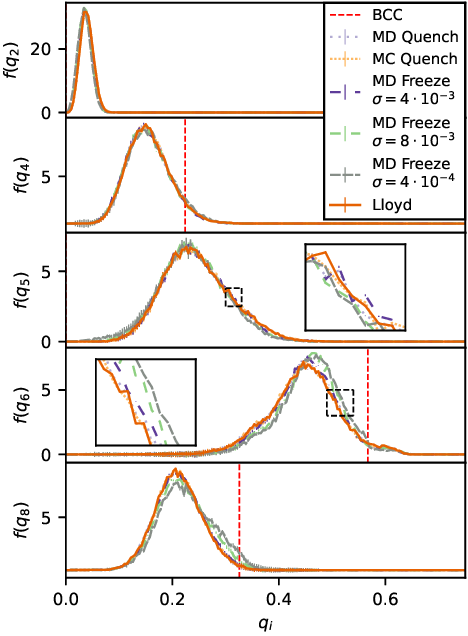}
  \caption{\label{fig:sphmink_quench} Local order and structure
    measures of final structure obtained by MD, MC and Lloyd quenches,
    as well as fast freezing MD simulations. Shown are the rotational
    invariants $q_{2,4,5,6,8}$ of the face-normal Minkowski
    tensors. These robust and sensitive measures characterise the
    shape of single Voronoi cells and thus local order and structure
    of the systems. The values of a BCC shaped cell is shown for
    reference. The data shows good agreement of the final structures
    of MD, MC and Lloyd quenches within statistical uncertainties
    indicating that the final structures of the MD and MC quenches are
    identical to the ones of the Lloyd quench. The fast freezing MD
    simulations show good agreement in $q_2$ and $q_4$, however,
    increasing differences in $q_5$,$q_6$ and $q_8$ with increasing
    cooling rate, indicating a differences in the local order compared
    to the structures obtained by the quenches. These findings are in
    line with the decreasing energies of the final structures with
    increasing cooling rates.}
\end{figure}

\begin{figure}
  \includegraphics{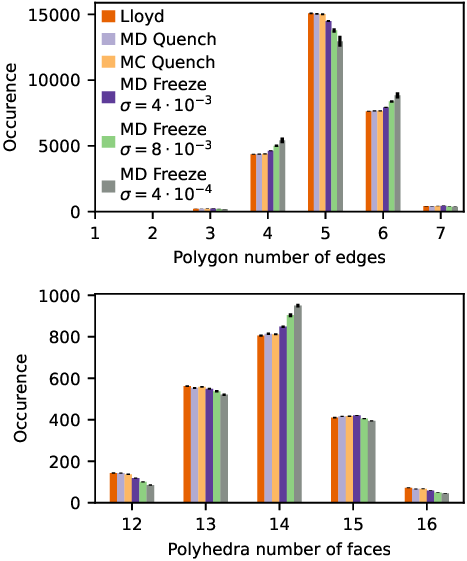}
  \caption{\label{fig:voronoi_stats_quench} Cell and face order
    distributions of the Voronoi cells in the structures obtained by
    MD, MC and Lloyd quenches as well as fast freezed MD
    simulations. Shown is the number of faces with $n$ edges (top) and
    the number of cells with $n$ faces in the Voronoi tessellation of
    the final structures. The data shows good agreement within
    statistical limits of the MD, MC and Lloyd quenches, where as the
    fast freezing MD runs show significant deviations increasing with
    increasing cooling rate.}
\end{figure}

Our results are shown in fig.~(\ref{fig:sphmink_quench}) and
fig.~(\ref{fig:voronoi_stats_quench}). We present the shape metrics
and Voronoi cell statistics for the final structures obtained MC, MD
and Lloyd quenches, as well as fast cooling MD simulations. For each
quench type the distribution is averaged over 24 individual simulation
runs. The data shows good agreement within statistical limits of the
distributions of the MD, MC and Lloyd quenches. These findings further
support the equality of the final structures, as previously indicated
by equal energies as well as $\tau$ values.

The distributions of the final structures obtained by quickly cooled
systems, however, show significant differences compared to the
quenched structures. A clear trend is visible: the higher the (absolute value of the)
cooling rate the higher the deviation from the inherent structure. This
is visible in the Minkowski structure metrics, but even more in the
Voronoi statistics. These results are in line with a similar trend in
the final energies, which decrease - hinting towards more optimal
structures - with increasing heating rates.

\subsection{Simulation parameters}
\label{sec:sim-params}

\textbf{Quenches, fig.~(\ref{fig:quench}), table~(\ref{tab:energies})}
All initial configurations are binomial point processes (ideal gases)
with $N=2000$ and $\rho=1$. The structure factors, $\tau$ and energy
values of MC, MD and Lloyd's algorithm runs are averaged over 24
individual runs. \textbf{MD:} MD step size chosen as $\Delta t=0.005$,
initial temperature set to $kT=2.1$. An initial set of 5000\,MD steps
are performed at $kT=2.1$ to equilibrate the initial configuration,
then the temperature is set to $kT=0$ and the system is run for $6
\times 10^{5}$\,MD steps. \textbf{MC:} Immediately after
initialization temperature is set to $kT=0$, then a total of $8.1
\times 10^{8}$\,MC steps are run. Step size is adapted every $3 \times
10^{7}$ steps. \textbf{Lloyd's algorithm:} A total of 50000 steps are
run. While the $\tau$ value in table~(\ref{tab:energies}) was taken
from the supplementary material from \cite{klatt_universal_2019}, the
structure factor shown in fig.~(\ref{fig:quench}) was generated by us
from the dataset
\textsc{3D-Final-configuration-derived-from-Binomial-PP-1.dat}
provided by \citet{klatt_universal_2019}. All structure factors are
computed with a bin width of 0.25 and a cutoff $k_{\text{max}}=25$.

\textbf{MD melting/freezing, fig.~(\ref{fig:md_phase_transition})} 48
individual runs for each melting and freezing were
run. \textbf{Melting} runs were initialized as BCC lattices, with each
component of the velocities randomly drawn from a normal distribution
to match a system temperature of $kT=0.1$. The velocities are modified
so the center of mass of the system is at rest. The initial thermostat
temperature is set to $kT=0.1$. \textbf{Freezing} runs were
initialized as ideal gas, thus each position component randomly chosen
to uniformly cover the simulation box with random velocities matching
an initial temperature of $kT=2.1$ with the center of mass of the
system at rest. For both melting and freezing runs the thermostat is
initialised with $Q=20$, $s=1$ and $\dot{s}=1$, the MD time step is
chosen as $\Delta t=0.001$. All systems ran a set of $5\times 10^{5}$
initial relax steps to equilibrate the system and thermostat at the
respective initial temperature. Then 800\,(melting)/840\,(freezing)
temperature steps, each with an increment of $\Delta kT= \pm 2.5
\times 10^{-3}$ are run, resulting in a cooling rate of
$\sigma=\frac{2.5 \times 10^{-3} \epsilon}{10000 \cdot 0.001\delta} =
2.5 \times 10^{-4} \epsilon / \delta$. For each temperature, 10000 MD
relax steps are run followed by a measurement phase comprising 1500 MD
steps where every 25 steps the energy and structure factor is
measured. For the isotherms 4 runs at each temperature for each
freezing and melting systems are run. The particles in melting systems
are initialized on a BCC lattice with velocities drawn from a normal
distribution to match their respective temperature with a resting
center of mass. The inital positions of the particles in the
isothermal freezing systems are uniformly distributed over the
simulation box with velocities drawn from a normal distribution to
match their respective temperature (ideal gas). Both melting and
freezing isothermal systems are then run for $8.5\cdot 10^{5}$ steps
with a time step of $\Delta t=0.001$. The thermostat is initialized
with $Q=20$, $s=1$ and $\dot{s}=1$. The mean energies are averaged
over the last $1.7\cdot 10^{5}$ steps.

\textbf{MC melting/freezing, fig.~(\ref{fig:mc_phase_transition})}
Since MD simulations outperforms MC code by far, we used MD
simulations to generate initial configurations very close to the phase
transition region and then continued these runs with MC code. The
parameters for the MD runs are identical to the ones mentioned above
with these exceptions: no initial relax steps are performed, the
positions of the particles in the systems initialized as ideal gas are
identical across all systems, however, do have randomly chosen
velocities. The in total 48 runs were given different initial
thermostat settings, where two runs shared shared one of the
combinations between $Q \in \{60, 50, 40, 30, 20, 10\}$ and $(s,
\dot{s}) \in \{(0.2, 0), (1, 0.5), (1.5, 1), (4, 3)\}$. Only 500
measure steps were performed at each temperature. The initial
configurations for the MC runs were taken after the relax phase at
$kT=1.87$\,(melting) and $kT=1.02$\,(freezing). A total of
213\,(melting)/320\,(freezing) temperature increments each with
$\Delta kT=\pm 4.7\times 10^{-4}$ were simulated. At each temperature
a total of $6\times 10^{6}$ MC relax steps were performed, followed by
$3.8 \times 10^{4}$ measurement steps, with $2000$ MC steps in between
individual measurements.

\textbf{Melting of the converged Lloyd states,
  fig.~(\ref{fig:klatt-melting})} The 48 individual MD simulations are
initialized with converged Lloyd states, thus $N=2000$ and
$\rho=1$. Random temperatures are assigned to match an initial
temperature $kT=0.05$ with the center of mass of the system at
rest. $2\times10^{5}$ initial relax steps are performed. Thermostat is
initialized with $Q=20$ and $s=\dot{s}=1$. MD time step is chosen as
$\Delta t=0.001$. A total of 840 temperature increments with $\Delta
kT=2.5 \times 10^{-3}$ are computed. Each temperature has 10000 relax
steps and 1500 measurements steps with 25 steps in between individual
measurements. All data represented by symbols are MD simulations
initialized as ideal gas, with random velocities matching their
initial temperatures, with the center of mass at rest. All ran a set
of initial relax steps of 20000 MD steps, had 2400 relax and 100
measurement MD steps for each temperature increment, with 25 MD steps
between each measurement. The olive circles (``$\sigma=-4 \cdot
10^{-3}$'') is combined data from 10 sets, each set consisting of 24
averaged runs, starting at different initial temperatures
$kT=(2.5,2.55,2.6,2.65,2.7,2.75,2.8,2.85,2.9,2.95)$. Each set had six
temperature increments of $\Delta kT=-0.5$. Combining the interlaced
sets yields the curve as shown. The same was done for the olive
triangles (``$\sigma=-8 \cdot 10^{-3}$''), however, only 2 sets (each
averaging over 24 runs) started at $kT=(2.1,2.15)$ and had 22
temperature increments with $\Delta kT=-0.1$. The olive pyramids
(``$\sigma=-4 \cdot 10^{-4}$'') is a single set averaging over 24 runs
starting at $kT=2.1$ with 43 temperature increments with $\Delta
kT=-0.05$.

\bibliography{QuanTizer}

\begin{thebibliography}{43}%
\makeatletter
\providecommand \@ifxundefined [1]{%
 \@ifx{#1\undefined}
}%
\providecommand \@ifnum [1]{%
 \ifnum #1\expandafter \@firstoftwo
 \else \expandafter \@secondoftwo
 \fi
}%
\providecommand \@ifx [1]{%
 \ifx #1\expandafter \@firstoftwo
 \else \expandafter \@secondoftwo
 \fi
}%
\providecommand \natexlab [1]{#1}%
\providecommand \enquote  [1]{``#1''}%
\providecommand \bibnamefont  [1]{#1}%
\providecommand \bibfnamefont [1]{#1}%
\providecommand \citenamefont [1]{#1}%
\providecommand \href@noop [0]{\@secondoftwo}%
\providecommand \href [0]{\begingroup \@sanitize@url \@href}%
\providecommand \@href[1]{\@@startlink{#1}\@@href}%
\providecommand \@@href[1]{\endgroup#1\@@endlink}%
\providecommand \@sanitize@url [0]{\catcode `\\12\catcode `\$12\catcode
  `\&12\catcode `\#12\catcode `\^12\catcode `\_12\catcode `\%12\relax}%
\providecommand \@@startlink[1]{}%
\providecommand \@@endlink[0]{}%
\providecommand \url  [0]{\begingroup\@sanitize@url \@url }%
\providecommand \@url [1]{\endgroup\@href {#1}{\urlprefix }}%
\providecommand \urlprefix  [0]{URL }%
\providecommand \Eprint [0]{\href }%
\providecommand \doibase [0]{http://dx.doi.org/}%
\providecommand \selectlanguage [0]{\@gobble}%
\providecommand \bibinfo  [0]{\@secondoftwo}%
\providecommand \bibfield  [0]{\@secondoftwo}%
\providecommand \translation [1]{[#1]}%
\providecommand \BibitemOpen [0]{}%
\providecommand \bibitemStop [0]{}%
\providecommand \bibitemNoStop [0]{.\EOS\space}%
\providecommand \EOS [0]{\spacefactor3000\relax}%
\providecommand \BibitemShut  [1]{\csname bibitem#1\endcsname}%
\let\auto@bib@innerbib\@empty
\bibitem [{\citenamefont {Weaire}\ and\ \citenamefont
  {Phelan}(1994)}]{weaire_counter-example_1994}%
  \BibitemOpen
  \bibfield  {author} {\bibinfo {author} {\bibfnamefont {D.}~\bibnamefont
  {Weaire}}\ and\ \bibinfo {author} {\bibfnamefont {R.}~\bibnamefont
  {Phelan}},\ }\href {\doibase 10.1080/09500839408241577} {\bibfield  {journal}
  {\bibinfo  {journal} {Philos. Mag. Lett.}\ }\textbf {\bibinfo {volume}
  {69}},\ \bibinfo {pages} {107} (\bibinfo {year} {1994})}\BibitemShut
  {NoStop}%
\bibitem [{\citenamefont {Weaire}(1997)}]{weaire_kelvin_1997}%
  \BibitemOpen
  \bibfield  {author} {\bibinfo {author} {\bibfnamefont {D.}~\bibnamefont
  {Weaire}},\ }\href@noop {} {\emph {\bibinfo {title} {The {{Kelvin
  Problem}}}}}\ (\bibinfo  {publisher} {{Taylor \& Francis}},\ \bibinfo
  {address} {{London}},\ \bibinfo {year} {1997})\BibitemShut {NoStop}%
\bibitem [{\citenamefont {Hales}(2005)}]{hales_proof_2005}%
  \BibitemOpen
  \bibfield  {author} {\bibinfo {author} {\bibfnamefont {T.}~\bibnamefont
  {Hales}},\ }\href {\doibase 10.4007/annals.2005.162.1065} {\bibfield
  {journal} {\bibinfo  {journal} {Ann. Math.}\ }\textbf {\bibinfo {volume}
  {162}},\ \bibinfo {pages} {1065} (\bibinfo {year} {2005})}\BibitemShut
  {NoStop}%
\bibitem [{\citenamefont {Torquato}, \citenamefont {Zhang},\ and\ \citenamefont
  {Stillinger}(2015)}]{torquato_ensemble_2015}%
  \BibitemOpen
  \bibfield  {author} {\bibinfo {author} {\bibfnamefont {S.}~\bibnamefont
  {Torquato}}, \bibinfo {author} {\bibfnamefont {G.}~\bibnamefont {Zhang}}, \
  and\ \bibinfo {author} {\bibfnamefont {F.~H.}\ \bibnamefont {Stillinger}},\
  }\href {\doibase 10.1103/PhysRevX.5.021020} {\bibfield  {journal} {\bibinfo
  {journal} {Phys. Rev. X}\ }\textbf {\bibinfo {volume} {5}},\ \bibinfo {pages}
  {021020} (\bibinfo {year} {2015})}\BibitemShut {NoStop}%
\bibitem [{\citenamefont {Zhang}, \citenamefont {Stillinger},\ and\
  \citenamefont {Torquato}(2015{\natexlab{a}})}]{zhang_ground_2015-1}%
  \BibitemOpen
  \bibfield  {author} {\bibinfo {author} {\bibfnamefont {G.}~\bibnamefont
  {Zhang}}, \bibinfo {author} {\bibfnamefont {F.~H.}\ \bibnamefont
  {Stillinger}}, \ and\ \bibinfo {author} {\bibfnamefont {S.}~\bibnamefont
  {Torquato}},\ }\href {\doibase 10.1103/PhysRevE.92.022119} {\bibfield
  {journal} {\bibinfo  {journal} {Phys. Rev. E}\ }\textbf {\bibinfo {volume}
  {92}},\ \bibinfo {pages} {022119} (\bibinfo {year}
  {2015}{\natexlab{a}})}\BibitemShut {NoStop}%
\bibitem [{\citenamefont {Zhang}, \citenamefont {Stillinger},\ and\
  \citenamefont {Torquato}(2015{\natexlab{b}})}]{zhang_ground_2015}%
  \BibitemOpen
  \bibfield  {author} {\bibinfo {author} {\bibfnamefont {G.}~\bibnamefont
  {Zhang}}, \bibinfo {author} {\bibfnamefont {F.~H.}\ \bibnamefont
  {Stillinger}}, \ and\ \bibinfo {author} {\bibfnamefont {S.}~\bibnamefont
  {Torquato}},\ }\href {\doibase 10.1103/PhysRevE.92.022120} {\bibfield
  {journal} {\bibinfo  {journal} {Phys. Rev. E}\ }\textbf {\bibinfo {volume}
  {92}},\ \bibinfo {pages} {022120} (\bibinfo {year}
  {2015}{\natexlab{b}})}\BibitemShut {NoStop}%
\bibitem [{\citenamefont {Zhang}, \citenamefont {Stillinger},\ and\
  \citenamefont {Torquato}(2017)}]{zhang_can_2017}%
  \BibitemOpen
  \bibfield  {author} {\bibinfo {author} {\bibfnamefont {G.}~\bibnamefont
  {Zhang}}, \bibinfo {author} {\bibfnamefont {F.~H.}\ \bibnamefont
  {Stillinger}}, \ and\ \bibinfo {author} {\bibfnamefont {S.}~\bibnamefont
  {Torquato}},\ }\href {\doibase 10.1039/C7SM01028A} {\bibfield  {journal}
  {\bibinfo  {journal} {Soft Matter}\ }\textbf {\bibinfo {volume} {13}},\
  \bibinfo {pages} {6197} (\bibinfo {year} {2017})}\BibitemShut {NoStop}%
\bibitem [{\citenamefont {Ghosh}\ and\ \citenamefont
  {Lebowitz}(2018)}]{ghosh_generalized_2018}%
  \BibitemOpen
  \bibfield  {author} {\bibinfo {author} {\bibfnamefont {S.}~\bibnamefont
  {Ghosh}}\ and\ \bibinfo {author} {\bibfnamefont {J.~L.}\ \bibnamefont
  {Lebowitz}},\ }\href {\doibase 10.1007/s00220-018-3226-5} {\bibfield
  {journal} {\bibinfo  {journal} {Commun. Math. Phys.}\ }\textbf {\bibinfo
  {volume} {363}},\ \bibinfo {pages} {97} (\bibinfo {year} {2018})}\BibitemShut
  {NoStop}%
\bibitem [{\citenamefont {Zhang}, \citenamefont {Stillinger},\ and\
  \citenamefont {Torquato}(2016)}]{zhang_perfect_2016}%
  \BibitemOpen
  \bibfield  {author} {\bibinfo {author} {\bibfnamefont {G.}~\bibnamefont
  {Zhang}}, \bibinfo {author} {\bibfnamefont {F.~H.}\ \bibnamefont
  {Stillinger}}, \ and\ \bibinfo {author} {\bibfnamefont {S.}~\bibnamefont
  {Torquato}},\ }\href {\doibase 10.1038/srep36963} {\bibfield  {journal}
  {\bibinfo  {journal} {Sci. Rep.}\ }\textbf {\bibinfo {volume} {6}},\ \bibinfo
  {pages} {36963} (\bibinfo {year} {2016})}\BibitemShut {NoStop}%
\bibitem [{\citenamefont {Bi}\ \emph {et~al.}(2015)\citenamefont {Bi},
  \citenamefont {Lopez}, \citenamefont {Schwarz},\ and\ \citenamefont
  {Manning}}]{bi_density-independent_2015}%
  \BibitemOpen
  \bibfield  {author} {\bibinfo {author} {\bibfnamefont {D.}~\bibnamefont
  {Bi}}, \bibinfo {author} {\bibfnamefont {J.~H.}\ \bibnamefont {Lopez}},
  \bibinfo {author} {\bibfnamefont {J.~M.}\ \bibnamefont {Schwarz}}, \ and\
  \bibinfo {author} {\bibfnamefont {M.~L.}\ \bibnamefont {Manning}},\ }\href
  {\doibase 10.1038/nphys3471} {\bibfield  {journal} {\bibinfo  {journal} {Nat.
  Phys.}\ }\textbf {\bibinfo {volume} {11}},\ \bibinfo {pages} {1074} (\bibinfo
  {year} {2015})}\BibitemShut {NoStop}%
\bibitem [{\citenamefont {Merkel}\ and\ \citenamefont
  {Manning}(2018)}]{merkel_geometrically_2018}%
  \BibitemOpen
  \bibfield  {author} {\bibinfo {author} {\bibfnamefont {M.}~\bibnamefont
  {Merkel}}\ and\ \bibinfo {author} {\bibfnamefont {M.~L.}\ \bibnamefont
  {Manning}},\ }\href {\doibase 10.1088/1367-2630/aaaa13} {\bibfield  {journal}
  {\bibinfo  {journal} {New J. Phys.}\ }\textbf {\bibinfo {volume} {20}},\
  \bibinfo {pages} {022002} (\bibinfo {year} {2018})}\BibitemShut {NoStop}%
\bibitem [{\citenamefont {Gersho}(1979)}]{gersho_asymptotically_1979}%
  \BibitemOpen
  \bibfield  {author} {\bibinfo {author} {\bibfnamefont {A.}~\bibnamefont
  {Gersho}},\ }\href@noop {} {\bibfield  {journal} {\bibinfo  {journal} {IEEE
  Trans. Inf. Theory}\ }\textbf {\bibinfo {volume} {25}},\ \bibinfo {pages}
  {373} (\bibinfo {year} {1979})}\BibitemShut {NoStop}%
\bibitem [{\citenamefont {Lloyd}(1982)}]{lloyd_least_1982}%
  \BibitemOpen
  \bibfield  {author} {\bibinfo {author} {\bibfnamefont {S.}~\bibnamefont
  {Lloyd}},\ }\href {\doibase 10.1109/TIT.1982.1056489} {\bibfield  {journal}
  {\bibinfo  {journal} {IEEE Trans. Inf. Theory}\ }\textbf {\bibinfo {volume}
  {28}},\ \bibinfo {pages} {129} (\bibinfo {year} {1982})}\BibitemShut
  {NoStop}%
\bibitem [{\citenamefont {Conway}\ and\ \citenamefont
  {Sloane}(1999)}]{conway_sphere_1999}%
  \BibitemOpen
  \bibfield  {author} {\bibinfo {author} {\bibfnamefont {J.}~\bibnamefont
  {Conway}}\ and\ \bibinfo {author} {\bibfnamefont {N.~J.~A.}\ \bibnamefont
  {Sloane}},\ }\href {\doibase 10.1007/978-1-4757-6568-7} {\emph {\bibinfo
  {title} {Sphere {{Packings}}, {{Lattices}} and {{Groups}}}}},\ \bibinfo
  {edition} {3rd}\ ed.,\ Grundlehren Der Mathematischen {{Wissenschaften}}\
  (\bibinfo  {publisher} {{Springer-Verlag}},\ \bibinfo {address} {{New
  York}},\ \bibinfo {year} {1999})\BibitemShut {NoStop}%
\bibitem [{\citenamefont {Okabe}\ \emph {et~al.}(2000)\citenamefont {Okabe},
  \citenamefont {Boots}, \citenamefont {Sugihara},\ and\ \citenamefont
  {Chiu}}]{okabe_spatial_2000}%
  \BibitemOpen
  \bibfield  {author} {\bibinfo {author} {\bibfnamefont {A.}~\bibnamefont
  {Okabe}}, \bibinfo {author} {\bibfnamefont {B.}~\bibnamefont {Boots}},
  \bibinfo {author} {\bibfnamefont {K.}~\bibnamefont {Sugihara}}, \ and\
  \bibinfo {author} {\bibfnamefont {S.~N.}\ \bibnamefont {Chiu}},\ }\href@noop
  {} {\emph {\bibinfo {title} {Spatial {{Tessellations}}: {{Concepts}} and
  {{Applications}} of {{Voronoi Diagrams}}}}},\ \bibinfo {edition} {2nd}\ ed.\
  (\bibinfo  {publisher} {{Wiley}},\ \bibinfo {address} {{Chichester; New
  York}},\ \bibinfo {year} {2000})\BibitemShut {NoStop}%
\bibitem [{\citenamefont {Du}\ and\ \citenamefont
  {Wang}(2005)}]{du_optimal_2005}%
  \BibitemOpen
  \bibfield  {author} {\bibinfo {author} {\bibfnamefont {Q.}~\bibnamefont
  {Du}}\ and\ \bibinfo {author} {\bibfnamefont {D.}~\bibnamefont {Wang}},\
  }\href {\doibase 10.1016/j.camwa.2004.12.008} {\bibfield  {journal} {\bibinfo
   {journal} {Comput. Math. Appl.}\ }\textbf {\bibinfo {volume} {49}},\
  \bibinfo {pages} {1355} (\bibinfo {year} {2005})}\BibitemShut {NoStop}%
\bibitem [{\citenamefont {Du}, \citenamefont {Gunzburger},\ and\ \citenamefont
  {Ju}(2010)}]{du_advances_2010}%
  \BibitemOpen
  \bibfield  {author} {\bibinfo {author} {\bibfnamefont {Q.}~\bibnamefont
  {Du}}, \bibinfo {author} {\bibfnamefont {M.}~\bibnamefont {Gunzburger}}, \
  and\ \bibinfo {author} {\bibfnamefont {L.}~\bibnamefont {Ju}},\ }\href
  {\doibase 10.4208/nmtma.2010.32s.1} {\bibfield  {journal} {\bibinfo
  {journal} {Numer Math Theor Meth Appl}\ }\textbf {\bibinfo {volume} {3}},\
  \bibinfo {pages} {119} (\bibinfo {year} {2010})}\BibitemShut {NoStop}%
\bibitem [{\citenamefont {Gray}(1984)}]{1162229}%
  \BibitemOpen
  \bibfield  {author} {\bibinfo {author} {\bibfnamefont {R.}~\bibnamefont
  {Gray}},\ }\href@noop {} {\bibfield  {journal} {\bibinfo  {journal} {IEEE
  ASSP Mag.}\ }\textbf {\bibinfo {volume} {1}},\ \bibinfo {pages} {4} (\bibinfo
  {year} {1984})}\BibitemShut {NoStop}%
\bibitem [{\citenamefont {Lu}\ \emph {et~al.}(2012)\citenamefont {Lu},
  \citenamefont {Sun}, \citenamefont {Pan},\ and\ \citenamefont
  {Wang}}]{6143938}%
  \BibitemOpen
  \bibfield  {author} {\bibinfo {author} {\bibfnamefont {L.}~\bibnamefont
  {Lu}}, \bibinfo {author} {\bibfnamefont {F.}~\bibnamefont {Sun}}, \bibinfo
  {author} {\bibfnamefont {H.}~\bibnamefont {Pan}}, \ and\ \bibinfo {author}
  {\bibfnamefont {W.}~\bibnamefont {Wang}},\ }\href@noop {} {\bibfield
  {journal} {\bibinfo  {journal} {IEEE Trans. Vis. Comput. Graph.}\ }\textbf
  {\bibinfo {volume} {18}},\ \bibinfo {pages} {1880} (\bibinfo {year}
  {2012})}\BibitemShut {NoStop}%
\bibitem [{\citenamefont {Torquato}(2010)}]{torquato_reformulation_2010}%
  \BibitemOpen
  \bibfield  {author} {\bibinfo {author} {\bibfnamefont {S.}~\bibnamefont
  {Torquato}},\ }\href {\doibase 10.1103/PhysRevE.82.056109} {\bibfield
  {journal} {\bibinfo  {journal} {Phys. Rev. E}\ }\textbf {\bibinfo {volume}
  {82}} (\bibinfo {year} {2010}),\ 10.1103/PhysRevE.82.056109}\BibitemShut
  {NoStop}%
\bibitem [{\citenamefont {Ruscher}, \citenamefont {Baschnagel},\ and\
  \citenamefont {Farago}(2015)}]{ruscher_voronoi_2015}%
  \BibitemOpen
  \bibfield  {author} {\bibinfo {author} {\bibfnamefont {C.}~\bibnamefont
  {Ruscher}}, \bibinfo {author} {\bibfnamefont {J.}~\bibnamefont {Baschnagel}},
  \ and\ \bibinfo {author} {\bibfnamefont {J.}~\bibnamefont {Farago}},\ }\href
  {\doibase 10.1209/0295-5075/112/66003} {\bibfield  {journal} {\bibinfo
  {journal} {EPL}\ }\textbf {\bibinfo {volume} {112}},\ \bibinfo {pages}
  {66003} (\bibinfo {year} {2015})}\BibitemShut {NoStop}%
\bibitem [{\citenamefont {Ruscher}(2017)}]{ruscher_voronoi_2017}%
  \BibitemOpen
  \bibfield  {author} {\bibinfo {author} {\bibfnamefont {C.}~\bibnamefont
  {Ruscher}},\ }\emph {\bibinfo {title} {The {{Voronoi}} Liquid : {{A}} New
  Model to Probe the Glass Transition}},\ \href@noop {} {\bibinfo {type}
  {{{PhD}} thesis}},\ \bibinfo  {school} {\'Ecole doctorale Physique et
  chimie-physique} (\bibinfo {year} {2017})\BibitemShut {NoStop}%
\bibitem [{\citenamefont {Klatt}\ \emph {et~al.}(2019)\citenamefont {Klatt},
  \citenamefont {Lovri{\'c}}, \citenamefont {Chen}, \citenamefont {Kapfer},
  \citenamefont {Schaller}, \citenamefont {Sch{\"o}nh{\"o}fer}, \citenamefont
  {Gardiner}, \citenamefont {Smith}, \citenamefont {{Schr{\"o}der-Turk}},\ and\
  \citenamefont {Torquato}}]{klatt_universal_2019}%
  \BibitemOpen
  \bibfield  {author} {\bibinfo {author} {\bibfnamefont {M.~A.}\ \bibnamefont
  {Klatt}}, \bibinfo {author} {\bibfnamefont {J.}~\bibnamefont {Lovri{\'c}}},
  \bibinfo {author} {\bibfnamefont {D.}~\bibnamefont {Chen}}, \bibinfo {author}
  {\bibfnamefont {S.~C.}\ \bibnamefont {Kapfer}}, \bibinfo {author}
  {\bibfnamefont {F.~M.}\ \bibnamefont {Schaller}}, \bibinfo {author}
  {\bibfnamefont {P.~W.~A.}\ \bibnamefont {Sch{\"o}nh{\"o}fer}}, \bibinfo
  {author} {\bibfnamefont {B.~S.}\ \bibnamefont {Gardiner}}, \bibinfo {author}
  {\bibfnamefont {A.-S.}\ \bibnamefont {Smith}}, \bibinfo {author}
  {\bibfnamefont {G.~E.}\ \bibnamefont {{Schr{\"o}der-Turk}}}, \ and\ \bibinfo
  {author} {\bibfnamefont {S.}~\bibnamefont {Torquato}},\ }\href {\doibase
  10.1038/s41467-019-08360-5} {\bibfield  {journal} {\bibinfo  {journal} {Nat.
  Commun.}\ }\textbf {\bibinfo {volume} {10}},\ \bibinfo {pages} {811}
  (\bibinfo {year} {2019})}\BibitemShut {NoStop}%
\bibitem [{\citenamefont {{Schr{\"o}der-Turk}}\ \emph
  {et~al.}(2011)\citenamefont {{Schr{\"o}der-Turk}}, \citenamefont {Mickel},
  \citenamefont {Kapfer}, \citenamefont {Klatt}, \citenamefont {Schaller},
  \citenamefont {Hoffmann}, \citenamefont {Kleppmann}, \citenamefont
  {Armstrong}, \citenamefont {Inayat}, \citenamefont {Hug}, \citenamefont
  {Reichelsdorfer}, \citenamefont {Peukert}, \citenamefont {Schwieger},\ and\
  \citenamefont {Mecke}}]{schroder-turk_minkowski_2011}%
  \BibitemOpen
  \bibfield  {author} {\bibinfo {author} {\bibfnamefont {G.~E.}\ \bibnamefont
  {{Schr{\"o}der-Turk}}}, \bibinfo {author} {\bibfnamefont {W.}~\bibnamefont
  {Mickel}}, \bibinfo {author} {\bibfnamefont {S.~C.}\ \bibnamefont {Kapfer}},
  \bibinfo {author} {\bibfnamefont {M.~A.}\ \bibnamefont {Klatt}}, \bibinfo
  {author} {\bibfnamefont {F.~M.}\ \bibnamefont {Schaller}}, \bibinfo {author}
  {\bibfnamefont {M.~J.~F.}\ \bibnamefont {Hoffmann}}, \bibinfo {author}
  {\bibfnamefont {N.}~\bibnamefont {Kleppmann}}, \bibinfo {author}
  {\bibfnamefont {P.}~\bibnamefont {Armstrong}}, \bibinfo {author}
  {\bibfnamefont {A.}~\bibnamefont {Inayat}}, \bibinfo {author} {\bibfnamefont
  {D.}~\bibnamefont {Hug}}, \bibinfo {author} {\bibfnamefont {M.}~\bibnamefont
  {Reichelsdorfer}}, \bibinfo {author} {\bibfnamefont {W.}~\bibnamefont
  {Peukert}}, \bibinfo {author} {\bibfnamefont {W.}~\bibnamefont {Schwieger}},
  \ and\ \bibinfo {author} {\bibfnamefont {K.}~\bibnamefont {Mecke}},\ }\href
  {\doibase 10.1002/adma.201100562} {\bibfield  {journal} {\bibinfo  {journal}
  {Adv. Mater.}\ }\textbf {\bibinfo {volume} {23}},\ \bibinfo {pages} {2535}
  (\bibinfo {year} {2011})}\BibitemShut {NoStop}%
\bibitem [{\citenamefont {Thomson}(1887)}]{thomson_division_1887}%
  \BibitemOpen
  \bibfield  {author} {\bibinfo {author} {\bibfnamefont {W.}~\bibnamefont
  {Thomson}},\ }\href {\doibase 10.1007/BF02612322} {\bibfield  {journal}
  {\bibinfo  {journal} {Acta Math.}\ }\textbf {\bibinfo {volume} {11}},\
  \bibinfo {pages} {121} (\bibinfo {year} {1887})}\BibitemShut {NoStop}%
\bibitem [{\citenamefont {Ruscher}, \citenamefont {Baschnagel},\ and\
  \citenamefont {Farago}(2018)}]{ruscher_voronoi_2018}%
  \BibitemOpen
  \bibfield  {author} {\bibinfo {author} {\bibfnamefont {C.}~\bibnamefont
  {Ruscher}}, \bibinfo {author} {\bibfnamefont {J.}~\bibnamefont {Baschnagel}},
  \ and\ \bibinfo {author} {\bibfnamefont {J.}~\bibnamefont {Farago}},\ }\href
  {\doibase 10.1103/PhysRevE.97.032132} {\bibfield  {journal} {\bibinfo
  {journal} {Phys. Rev. E}\ }\textbf {\bibinfo {volume} {97}},\ \bibinfo
  {pages} {032132} (\bibinfo {year} {2018})}\BibitemShut {NoStop}%
\bibitem [{\citenamefont {Ruscher}\ \emph {et~al.}(2017)\citenamefont
  {Ruscher}, \citenamefont {Semenov}, \citenamefont {Baschnagel},\ and\
  \citenamefont {Farago}}]{ruscher_anomalous_2017}%
  \BibitemOpen
  \bibfield  {author} {\bibinfo {author} {\bibfnamefont {C.}~\bibnamefont
  {Ruscher}}, \bibinfo {author} {\bibfnamefont {A.~N.}\ \bibnamefont
  {Semenov}}, \bibinfo {author} {\bibfnamefont {J.}~\bibnamefont {Baschnagel}},
  \ and\ \bibinfo {author} {\bibfnamefont {J.}~\bibnamefont {Farago}},\ }\href
  {\doibase 10.1063/1.4979720} {\bibfield  {journal} {\bibinfo  {journal} {J.
  Chem. Phys.}\ }\textbf {\bibinfo {volume} {146}},\ \bibinfo {pages} {144502}
  (\bibinfo {year} {2017})}\BibitemShut {NoStop}%
\bibitem [{\citenamefont {Ruscher}\ \emph {et~al.}(2020)\citenamefont
  {Ruscher}, \citenamefont {Ciarella}, \citenamefont {Luo}, \citenamefont
  {Janssen}, \citenamefont {Farago},\ and\ \citenamefont
  {Baschnagel}}]{ruscher_glassy_2020}%
  \BibitemOpen
  \bibfield  {author} {\bibinfo {author} {\bibfnamefont {C.}~\bibnamefont
  {Ruscher}}, \bibinfo {author} {\bibfnamefont {S.}~\bibnamefont {Ciarella}},
  \bibinfo {author} {\bibfnamefont {C.}~\bibnamefont {Luo}}, \bibinfo {author}
  {\bibfnamefont {L.~M.~C.}\ \bibnamefont {Janssen}}, \bibinfo {author}
  {\bibfnamefont {J.}~\bibnamefont {Farago}}, \ and\ \bibinfo {author}
  {\bibfnamefont {J.}~\bibnamefont {Baschnagel}},\ }\href {\doibase
  10.1088/1361-648X/abc4cc} {\bibfield  {journal} {\bibinfo  {journal} {J.
  Phys.: Condens. Matter}\ }\textbf {\bibinfo {volume} {33}},\ \bibinfo {pages}
  {064001} (\bibinfo {year} {2020})}\BibitemShut {NoStop}%
\bibitem [{\citenamefont {Torquato}\ and\ \citenamefont
  {Stillinger}(2003)}]{torquato_local_2003}%
  \BibitemOpen
  \bibfield  {author} {\bibinfo {author} {\bibfnamefont {S.}~\bibnamefont
  {Torquato}}\ and\ \bibinfo {author} {\bibfnamefont {F.~H.}\ \bibnamefont
  {Stillinger}},\ }\href {\doibase 10.1103/PhysRevE.68.041113} {\bibfield
  {journal} {\bibinfo  {journal} {Phys. Rev. E}\ }\textbf {\bibinfo {volume}
  {68}},\ \bibinfo {pages} {041113} (\bibinfo {year} {2003})}\BibitemShut
  {NoStop}%
\bibitem [{\citenamefont {Ghosh}\ and\ \citenamefont
  {Lebowitz}(2017)}]{ghosh_fluctuations_2017}%
  \BibitemOpen
  \bibfield  {author} {\bibinfo {author} {\bibfnamefont {S.}~\bibnamefont
  {Ghosh}}\ and\ \bibinfo {author} {\bibfnamefont {J.~L.}\ \bibnamefont
  {Lebowitz}},\ }\href {\doibase 10.1007/s13226-017-0248-1} {\bibfield
  {journal} {\bibinfo  {journal} {Indian J. Pure Appl. Math.}\ }\textbf
  {\bibinfo {volume} {48}},\ \bibinfo {pages} {609} (\bibinfo {year}
  {2017})}\BibitemShut {NoStop}%
\bibitem [{\citenamefont {Torquato}(2018)}]{torquato_hyperuniform_2018}%
  \BibitemOpen
  \bibfield  {author} {\bibinfo {author} {\bibfnamefont {S.}~\bibnamefont
  {Torquato}},\ }\href {\doibase 10.1016/j.physrep.2018.03.001} {\bibfield
  {journal} {\bibinfo  {journal} {Phys. Rep.}\ }\textbf {\bibinfo {volume}
  {745}},\ \bibinfo {pages} {1} (\bibinfo {year} {2018})}\BibitemShut {NoStop}%
\bibitem [{\citenamefont {Atkinson}\ \emph {et~al.}(2016)\citenamefont
  {Atkinson}, \citenamefont {Zhang}, \citenamefont {Hopkins},\ and\
  \citenamefont {Torquato}}]{atkinson_critical_2016}%
  \BibitemOpen
  \bibfield  {author} {\bibinfo {author} {\bibfnamefont {S.}~\bibnamefont
  {Atkinson}}, \bibinfo {author} {\bibfnamefont {G.}~\bibnamefont {Zhang}},
  \bibinfo {author} {\bibfnamefont {A.~B.}\ \bibnamefont {Hopkins}}, \ and\
  \bibinfo {author} {\bibfnamefont {S.}~\bibnamefont {Torquato}},\ }\href
  {\doibase 10.1103/PhysRevE.94.012902} {\bibfield  {journal} {\bibinfo
  {journal} {Phys. Rev. E}\ }\textbf {\bibinfo {volume} {94}},\ \bibinfo
  {pages} {012902} (\bibinfo {year} {2016})}\BibitemShut {NoStop}%
\bibitem [{\citenamefont {Metropolis}\ \emph {et~al.}(1953)\citenamefont
  {Metropolis}, \citenamefont {Rosenbluth}, \citenamefont {Rosenbluth},
  \citenamefont {Teller},\ and\ \citenamefont
  {Teller}}]{metropolis_equation_1953}%
  \BibitemOpen
  \bibfield  {author} {\bibinfo {author} {\bibfnamefont {N.}~\bibnamefont
  {Metropolis}}, \bibinfo {author} {\bibfnamefont {A.~W.}\ \bibnamefont
  {Rosenbluth}}, \bibinfo {author} {\bibfnamefont {M.~N.}\ \bibnamefont
  {Rosenbluth}}, \bibinfo {author} {\bibfnamefont {A.~H.}\ \bibnamefont
  {Teller}}, \ and\ \bibinfo {author} {\bibfnamefont {E.}~\bibnamefont
  {Teller}},\ }\href {\doibase 10.1063/1.1699114} {\bibfield  {journal}
  {\bibinfo  {journal} {J. Chem. Phys.}\ }\textbf {\bibinfo {volume} {21}},\
  \bibinfo {pages} {1087} (\bibinfo {year} {1953})}\BibitemShut {NoStop}%
\bibitem [{\citenamefont {Kr{\"u}ger}\ and\ \citenamefont
  {Knauf}(2016)}]{kruger_mocasinns_2016}%
  \BibitemOpen
  \bibfield  {author} {\bibinfo {author} {\bibfnamefont {B.}~\bibnamefont
  {Kr{\"u}ger}}\ and\ \bibinfo {author} {\bibfnamefont {J.~F.}\ \bibnamefont
  {Knauf}},\ }\href@noop {} {\enquote {\bibinfo {title} {Mocasinns},}\ }
  (\bibinfo {year} {2016})\BibitemShut {NoStop}%
\bibitem [{\citenamefont {Kr{\"u}ger}(2016)}]{Krueger2016}%
  \BibitemOpen
  \bibfield  {author} {\bibinfo {author} {\bibfnamefont {B.}~\bibnamefont
  {Kr{\"u}ger}},\ }\emph {\bibinfo {title} {Simulating Triangulations:
  {{Graphs}}, Manifolds and (Quantum) Spacetime}},\ \href@noop {} {\bibinfo
  {type} {Doctoralthesis}},\ \bibinfo  {school} {FAU University Press}
  (\bibinfo {year} {2016})\BibitemShut {NoStop}%
\bibitem [{\citenamefont {{Schr{\"o}der-Turk}}\ \emph
  {et~al.}(2013)\citenamefont {{Schr{\"o}der-Turk}}, \citenamefont {Mickel},
  \citenamefont {Kapfer}, \citenamefont {Schaller}, \citenamefont
  {Breidenbach}, \citenamefont {Hug},\ and\ \citenamefont
  {Mecke}}]{schroder-turk_minkowski_2013}%
  \BibitemOpen
  \bibfield  {author} {\bibinfo {author} {\bibfnamefont {G.~E.}\ \bibnamefont
  {{Schr{\"o}der-Turk}}}, \bibinfo {author} {\bibfnamefont {W.}~\bibnamefont
  {Mickel}}, \bibinfo {author} {\bibfnamefont {S.~C.}\ \bibnamefont {Kapfer}},
  \bibinfo {author} {\bibfnamefont {F.~M.}\ \bibnamefont {Schaller}}, \bibinfo
  {author} {\bibfnamefont {B.}~\bibnamefont {Breidenbach}}, \bibinfo {author}
  {\bibfnamefont {D.}~\bibnamefont {Hug}}, \ and\ \bibinfo {author}
  {\bibfnamefont {K.}~\bibnamefont {Mecke}},\ }\href {\doibase
  10.1088/1367-2630/15/8/083028} {\bibfield  {journal} {\bibinfo  {journal}
  {New J. Phys.}\ }\textbf {\bibinfo {volume} {15}},\ \bibinfo {pages} {083028}
  (\bibinfo {year} {2013})}\BibitemShut {NoStop}%
\bibitem [{\citenamefont {Rycroft}(2009)}]{rycroft_voro_2009-1}%
  \BibitemOpen
  \bibfield  {author} {\bibinfo {author} {\bibfnamefont {C.~H.}\ \bibnamefont
  {Rycroft}},\ }\href {\doibase 10.1063/1.3215722} {\bibfield  {journal}
  {\bibinfo  {journal} {Chaos}\ }\textbf {\bibinfo {volume} {19}},\ \bibinfo
  {pages} {041111} (\bibinfo {year} {2009})}\BibitemShut {NoStop}%
\bibitem [{\citenamefont {Nos{\'e}}(1983)}]{nose_molecular_1983}%
  \BibitemOpen
  \bibfield  {author} {\bibinfo {author} {\bibfnamefont {S.}~\bibnamefont
  {Nos{\'e}}},\ }\href {\doibase 10.1080/00268970110089108} {\bibfield
  {journal} {\bibinfo  {journal} {Mol. Phys.}\ }\textbf {\bibinfo {volume}
  {100}},\ \bibinfo {pages} {191} (\bibinfo {year} {1983})}\BibitemShut
  {NoStop}%
\bibitem [{\citenamefont {Martyna}, \citenamefont {Tobias},\ and\ \citenamefont
  {Klein}(1994)}]{martyna_constant_1994}%
  \BibitemOpen
  \bibfield  {author} {\bibinfo {author} {\bibfnamefont {G.~J.}\ \bibnamefont
  {Martyna}}, \bibinfo {author} {\bibfnamefont {D.~J.}\ \bibnamefont {Tobias}},
  \ and\ \bibinfo {author} {\bibfnamefont {M.~L.}\ \bibnamefont {Klein}},\
  }\href {\doibase 10.1063/1.467468} {\bibfield  {journal} {\bibinfo  {journal}
  {J. Chem. Phys.}\ }\textbf {\bibinfo {volume} {101}},\ \bibinfo {pages}
  {4177} (\bibinfo {year} {1994})}\BibitemShut {NoStop}%
\bibitem [{\citenamefont {Hansen}\ and\ \citenamefont
  {McDonald}(2013)}]{hansen_theory_2013}%
  \BibitemOpen
  \bibfield  {author} {\bibinfo {author} {\bibfnamefont {J.-P.}\ \bibnamefont
  {Hansen}}\ and\ \bibinfo {author} {\bibfnamefont {I.~R.}\ \bibnamefont
  {McDonald}},\ }\href@noop {} {\emph {\bibinfo {title} {Theory of {{Simple
  Liquids}}: {{With Applications}} to {{Soft Matter}}}}},\ \bibinfo {edition}
  {4th}\ ed.\ (\bibinfo  {publisher} {{Academic Press}},\ \bibinfo {address}
  {{Amsterdam; Boston}},\ \bibinfo {year} {2013})\BibitemShut {NoStop}%
\bibitem [{Note1()}]{Note1}%
  \BibitemOpen
  \bibinfo {note} {We estimate $H$ by fitting a Gaussian-shaped density to the
  peak ($6<k<7.5$) and by extrapolating a second-degree polynomial $ax^2+c$ to
  the origin ($k_{\protect \qopname \relax m{min}}<k<4$). The precise value of
  $H$ depends on the fit range and functional form.}\BibitemShut {Stop}%
\bibitem [{\citenamefont {Wang}\ and\ \citenamefont
  {Landau}(2001)}]{wang_efficient_2001}%
  \BibitemOpen
  \bibfield  {author} {\bibinfo {author} {\bibfnamefont {F.}~\bibnamefont
  {Wang}}\ and\ \bibinfo {author} {\bibfnamefont {D.~P.}\ \bibnamefont
  {Landau}},\ }\href {\doibase 10.1103/PhysRevLett.86.2050} {\bibfield
  {journal} {\bibinfo  {journal} {Phys. Rev. Lett.}\ }\textbf {\bibinfo
  {volume} {86}},\ \bibinfo {pages} {2050} (\bibinfo {year}
  {2001})}\BibitemShut {NoStop}%
\bibitem [{Note2()}]{Note2}%
  \BibitemOpen
  \bibinfo {note} {See https://www.gnu.org/software/gsl/ for details on the
  implementation of the minimization methods and further
  information}\BibitemShut {NoStop}%
\end{thebibliography}%

\end{document}